%% file: pap_code.tex
\documentclass[aps,prb,preprint,showpacs,superscriptaddress,
floatfix,preprintnumbers,amsmath,amssymb]{revtex4}

\usepackage{comment}
\usepackage{graphicx}
\usepackage{dcolumn}
\usepackage{bm}
\usepackage{ifthen}
\usepackage{booktabs}
\usepackage{threeparttable}
\usepackage{multirow}
\usepackage{longtable}

\usepackage{color}
\usepackage[normalem]{ulem}

\newcommand{\CodeName}{\textsc{ABACUS}}

\newcommand{\Or}{\mathcal{O}}
\newcommand{\ie}{\textit{i.e.}{}}
\renewcommand{\Im}{\mathrm{Im}~}
\newcommand{\Tr}{\mathrm{Tr}}

\begin{document}

\title{Large-scale  {\it ab initio} simulations based on systematically improvable atomic basis}
\author{Pengfei Li}
\affiliation{Key Laboratory of Quantum Information, University of Science and Technology of China, Hefei, 230026, China}
\affiliation{Synergetic Innovation Center of Quantum Information and Quantum Physics, University of Science and Technology of China, Hefei, 230026, China}
\author{Xiaohui Liu}
\affiliation{Key Laboratory of Quantum Information, University of Science and Technology of China, Hefei, 230026, China}
\affiliation{Synergetic Innovation Center of Quantum Information and Quantum Physics, University of Science and Technology of China, Hefei, 230026, China}
\author{Mohan Chen}
\email{mohan.chen.chen.mohan@gmail.com}
\affiliation{Department of Mechanical and Aerospace Engineering, Princeton University, Princeton, New Jersey 08544, USA}
\author{Peize Lin}
\affiliation{Key Laboratory of Quantum Information, University of Science and Technology of China, Hefei, 230026, China}
\affiliation{Synergetic Innovation Center of Quantum Information and Quantum Physics, University of Science and Technology of China, Hefei, 230026, China}
\author{Xinguo Ren}
\email{renxg@ustc.edu.cn}
\affiliation{Key Laboratory of Quantum Information, University of Science and Technology of China, Hefei, 230026, China}
\affiliation{Synergetic Innovation Center of Quantum Information and Quantum Physics, University of Science and Technology of China, Hefei, 230026, China}
\author{Lin Lin}
\email{linlin@math.berkeley.edu}
\affiliation{
Department of Mathematics, University of California, Berkeley
and Computational Research Division, Lawrence Berkeley National
Laboratory, Berkeley, CA 94720, USA}
\author{Chao Yang}
\email{cyang@lbl.gov}
\affiliation{Computational Research Division, Lawrence Berkeley National Laboratory, Berkeley, CA 94720, USA}
\author{Lixin He}
\email{helx@ustc.edu.cn}
\affiliation{Key Laboratory of Quantum Information, University of Science and Technology of China, Hefei, 230026, China}
\affiliation{Synergetic Innovation Center of Quantum Information and Quantum Physics, University of Science and Technology of China, Hefei, 230026, China}

\date{\today }

\begin{abstract}
We present a first-principles computer code package (\CodeName) that is based on density functional theory and numerical atomic basis sets.
Theoretical foundations and numerical techniques used in the code are described,
with focus on the accuracy and transferability of the hierarchial atomic basis sets as generated using a scheme proposed by
Chen, Guo and He [J. Phys.:Condens. Matter \textbf{22}, 445501 (2010)].
Benchmark results are presented for a variety of systems include molecules, solids, surfaces, and defects.
All results show that the \CodeName~ package with its associated atomic basis sets
is an efficient and reliable tool for simulating both small and large-scale materials.
\end{abstract}

\maketitle
\include{introduction}
\include{method}
\include{result}

\include{summary}

\acknowledgments
The authors thank Yonghua Zhao and Wei Zhao for the valuable
help on the HPSEPS package. LH acknowledges the support from the Chinese National
Fundamental Research Program 2011CB921200, the
National Natural Science Funds for Distinguished Young Scholars. XR acknowledges
the support from Chinese National Science Foundation award number 11374276.
\begin{appendix}
\include{appendix}
\end{appendix}


\end{document}

%% file: introduction.tex
\section{\label{sec:intro}Introduction}

The density functional theory \cite{hohenberg64,kohn65} (DFT) based first-principles methods
are becoming increasingly important in the research fields of
condensed matter physics, material sciences, chemistry, and biology.
With the rapid development of supercomputers and the advances of numerical algorithms,
nowadays it is possible to study the electronic, structural and dynamical properties of
complicated physical systems containing thousands of atoms using DFT.
In these cases, the efficiency of widely used plane wave (PW) basis is largely limited,
because of its extended nature.
Instead, local bases, such as atomic orbitals, are the better choices.

Atomic orbitals have several advantages as basis sets
for the {\it ab initio} electronic structure calculations in the Kohn-Sham scheme. \cite{hohenberg64,kohn65}
First, the basis size of atomic orbitals is much smaller compared to other basis
sets, such as PW or real-space mesh.
Second, the atomic orbitals are strictly localized and therefore can be combined with either
the so-called linear scaling algorithms \cite{goedecker99} for electronic calculations,
or any other algorithm with a better scaling behavior than $O(N^3)$.
For example, Lin  {\it et al.} have recently developed a so-called Pole EXpansion and
Selected Inversion (PEXSI) technique, \cite{lin09,lin13} which takes
advantage of the sparsity of the Hamiltonian and the overlap
matrices obtained with local orbitals, and allows to solve the
Kohn-Sham equations with numerical effort that scales as $N^{\alpha}$
($\alpha \leq 2$) for both insulating and metallic systems, with $N$ being the number of atoms.


While the analytical Gaussian-type orbitals have been well established for
{\it ab initio} calculations in the quantum chemistry community for decades,
the numerically tabulated atomic orbitals are getting more and more popular in the
computational physics community.  Several first-principles codes
based on the numerical atomic orbitals have been developed in recent years,
{\it e.g.}, SIESTA, \cite{soler02} OpenMX, \cite{ozaki03} FHI-aims, \cite{volker08}
to name just a few, which aim at large-scale DFT calculations by exploiting the compactness
and locality of numerical atomic orbitals.
However, the numerical atomic orbitals must be constructed very carefully to
ensure both good accuracy and transferability. Furthermore, it would be
highly desirable if the quality of the basis sets can be systematically improved in an
unbiased way.
Recently, some of us [Chen, Guo, and He (CGH)] proposed a new scheme \cite{chen10,chen11} to construct
systematically improvable optimized atomic basis sets for DFT calculations.
Based on the CGH procedure for basis set generations,
we have developed a DFT package from scratch, named \emph{Atomic-orbital Based Ab-initio
Computation at UStc} (\CodeName)
here in the Key Laboratory of Quantum Information,
University of Science and Technology of China (USTC). In the \CodeName~ package, besides the primary option of using
numerical atomic orbitals as basis functions, PW basis can also be employed as an alternative choice.
This dual basis feature
is very useful for accuracy and consistency checks in benchmark calculations.
For both basis set choices, the package uses normal conserving pseudopotential in the
Unified Pseudopotential Format (UPF) 
that has been used in Quantum ESPRESSO.~\cite{qe2009}
The UPF pseudopotentials can be generated from the Opium package.\cite{OPIUM1}
Regarding the exchange-correlation functionals, we have implemented the local (spin) density approximation
[L(S)DA], and the generalized gradient approximation (GGA) as constructed by Perdew, Burke, and
Ernzerhof (PBE). In addition, semi-empirical van der Waals (vdW) corrected
DFT scheme as proposed by Grimme (DFT-D2) \cite{Grimme:2006b} has also been implemented.
Other advanced functionals such as hybrid functionals are currently under
development and will be reported in a later work.
At the level of LDA and GGA, the \CodeName~ package can do typical electronic structure calculations,
structure relaxations, and molecular dynamics.

In this paper, we first describe the main features of the \CodeName~ package,
as well as the major techniques that are used to implement DFT algorithms with atomic basis sets.
In a previous study,\cite{chen10} the CGH orbitals have been demonstrated to be accurate and
transferable for the group IV and group III-V semiconductors.
Here, we extend the tested systems to a larger range of elements,
including the alkali elements, 3d transition metals, group VI and group VII elements,
with focus on the structural and electronic properties of molecules, solids, surfaces, and defects.
The results demonstrate that \CodeName~ with the CGH orbitals are highly reliable
for both finite and extended systems. In particular, the
basis set at the level of double-$\zeta$ plus polarization function (DZP) is an excellent
choice to compromise between accuracy and computational cost,
and can be safely used in production calculations in most situations.

The rest of paper is organized as follows.
In Sec.~\ref{sec:methods}, we introduce the basic algorithms and numerical techniques.
In Sec.~\ref{sec:results}, we will demonstrate the performance of the \CodeName~ package,
focus on the accuracy of the atomic orbitals generated using the CGH scheme,
for a variety of benchmark systems.
Finally, we summarize our work in Sec. IV.

%% file: method.tex
\section{\label{sec:methods}Methods}

In this section, we first briefly recapitulate the basic formulation of solving Kohn-Sham equations in atomic basis (Sec.~\ref{sec:KSDFT})
to set up the stage.
This is followed by a description of the main techniques used in \CodeName.
Topics to be covered include the generation of the CGH atomic orbitals (Sec.~\ref{sec:basis_set}),
the construction of Hamiltonian matrix and overlap matrix (Sec.~\ref{sec:H}),
the solvers for Kohn-Sham equations (Sec.~\ref{sec:solver}),
and finally the total energy and force calculations (Sec.~\ref{sec:energy_force}).

\subsection{The Kohn-Sham equation in atomic basis}
\label{sec:KSDFT}
The central task in DFT calculations
is to solve the Kohn-Sham equation,\cite{hohenberg64,kohn65}
\begin{equation}
 {\hat H}_\text{KS}\Psi_n({\bf r})
 = \epsilon_n \Psi_n({\bf r}) \, ,
   \label{eq:Kohn_Sham}
\end{equation}
where $\epsilon_n$ and $\Psi_n(\bf r)$ are the Kohn-Sham eigenvalues and eigenfunctions for state $n$.
Hartree atomic unit ($e=\hbar=m_e=1$) is used here and throughtout the paper.
The Kohn-Sham Hamiltonian $\hat{H}_\text{KS}$ can be written as,
\begin{equation}
{\hat H}_\text{KS}=\hat{T}+\hat{V}^{\rm ext}({\bf r})+\hat{V}^{H}[\rho({\bf r})]+\hat{V}^{xc}[\rho({\bf r})]\, ,
 \label{eq:KS_hamil}
\end{equation}
where $\hat{T}=-{1 \over 2} \nabla^2$, $\hat{V}^{\rm ext}({\bf r})$, $\hat{V}^{H}[\rho({\bf r})]$, and
$\hat{V}^{xc}[\rho({\bf r})]$ are the kinetic energy operator, the external potential,
the Hartree potential, and the exchange-correlation
potential, respectively. The Kohn-Sham Hamiltonian ${\hat H}_\text{KS}$ thus depends on the
electron density $\rho(\bf r)$, which can be determined from the occupied Kohn-Sham orbitals
 \begin{equation}
    \rho({\bf r}) = 2 \sum_{n=1}^\text{occ.} |\Psi_n(\bf r)|^2\, .
  \label{eq:density}
 \end{equation}
Here for simplicity we assume that the system is spin-degenerate, and hence the spin index is omitted.
Extending the algorithm described here to the spin-polarized case is straightforward and
has been implemented in \CodeName.

Norm-conserving pseudopotentials are used to describe the ion-electron interactions.
The external potential $\hat{V}^{\rm ext}({\bf r})$ in Eq.~(\ref{eq:KS_hamil}) contains
the summation of the ion-electron potentials of all atoms plus, when they exist, applied external potentials.
Therefore (in the absence of the applied external potential),
\begin{equation}
\hat{V}^{\rm ext}({\bf r})=\sum_{{\bf R}}\sum_{\alpha i}\hat{v}_{\alpha}^{ps}({\bf r}-{\bf \tau}_{\alpha i}-{\bf R}) ,
\end{equation}
where $\hat{v}^{ps}_{\alpha i}$ is a norm-conserving pseudopotential \cite{hamann79} for the $i$-th atom of element
type $\alpha$, and ${\bf \tau}_{\alpha i}$ is the atomic coordinate in the cell ${\bf R}$.
The pseudopotential can split into a local part of the potential
$\hat{v}_{\alpha}^{L}$ and separable fully non-local potentials
\cite{kleinman82}
$\hat{v}^{NL}_{\alpha}$,
\begin{equation}
\hat{v}^{ps}_{\alpha}=\hat{v}_{\alpha}^{L}
+\hat{v}^{NL}_{\alpha}\, .
\end{equation}
The applied external potentials, {\it e.g.}, electric fields,
can be easily added to the local part of the potential,
while the non-local pseudopotential can be written as,
\begin{equation}
\hat{v}_{\alpha}^{NL}=\sum_{l=0}^{l_{max}}\sum_{m=-l}^{l}\sum_{n=1}^{n_{max}}
|\chi_{\alpha l m n}\rangle\langle\chi_{\alpha l m n}|,
\label{eq:nl_pp}
\end{equation}
where $|\chi_{\alpha lmn}\rangle$ are non-local projectors,
with $l$, $m$, $n$ being the angular momentum,
the magnetic momentum, and the multiplicity of projectors, respectively.
In Eq.~(\ref{eq:nl_pp}), $l_{max}$ and $n_{max}$ are the maximal angular
momentum and the maximal multiplicity of projectors for each angular momentum channel, respectively.

The Kohn-Sham equation is usually solved within certain basis sets.
The \CodeName~ package offers two choices of basis sets:
the PW basis set and the atomic basis set.
The advantage to do so is that the results obtained using atomic basis sets can be directly compared
to those obtained from PW basis sets for small systems, and thus provides valuable benchmarks for
the former. This will be clearly seen in Sec.~\ref{sec:results} where the benchmark results for
a variety of systems are presented.  However, since the PW algorithm has been well developed
and documented, here we only focus on the algorithms of the atomic-basis implementation.

Without losing generality, we consider crystalline systems under periodic boundary conditions.
The Kohn-Sham eigenfunctions in Eq.~(\ref{eq:Kohn_Sham}) then become Bloch orbitals which, within
atom-centered basis set, can be expanded as,
\begin{equation}
\Psi_{n\mathbf{k}}(\mathbf{r})={1 \over \sqrt{N}}
\sum_{\mathbf{R}}\sum_{\mu}
c_{n\mu,\mathbf{k}}e^{i\mathbf{k}\cdot\mathbf{R}}
\phi_{\mu}(\mathbf{r}-{\bf \tau}_{\alpha i} -\mathbf{R}),
\label{eq:Bloch_summation}
\end{equation}
where $\phi_{\mu}(\mathbf{r}-{\bf \tau}_{\alpha i} -\mathbf{R})$
are the atomic orbitals centering on the $i$-th atom of type $\alpha$ in the unit cell $\mathbf{R}$.
The orbital index $\mu$ is a compact one, $\mu = \{\alpha, i, l, m, \zeta \}$ with $l$ being the
angular momentum, $m$ the magnetic quantum number, and $\zeta$ the number of atomic orbitals for a given $l$.
Here $n$ and ${\bf k}$ are the band index and Bloch wave vector, and
$c_{n\mu,\mathbf{k}}$ are the Kohn-Sham eigen-coefficients. Finally $N$ is the number of unit cells in the
Born-von-Karmen supercell under the periodic boundary conditions.
Using Eq.~(\ref{eq:Bloch_summation}), the electron density within atom-centered basis sets can be computed as
\begin{eqnarray}
\label{eq.charge}
\rho(\mathbf{r})&=&\frac{1}{N_{k}}\sum_{n{\bf k}}
f_{n\mathbf{k}}
\Psi^{\ast}_{n\mathbf{k}}({\bf r})
\Psi_{n{\bf k}}({\bf r}) \nonumber \\
&=&\frac{1}{N_{k}}\sum_{\mathbf{R}}\sum_{\mu\nu}
\sum_{n{\bf k}}f_{n{\bf k}}
c_{\mu n,{\bf k}}^{\ast}
c_{n\nu,{\bf k}}
e^{-i\mathbf{k}\cdot \mathbf{R}}
\phi_{\mu}^{\ast}({\bf r}-\tau_{\alpha i}-\mathbf{R})
\phi_{\nu}({\bf r}-\tau_{\beta j})\nonumber \\
&=&\sum_{\mathbf{R}}\sum_{\mu\nu}\rho_{\mu\nu}(\mathbf{R})
\phi_{\mu}({\bf r}-\tau_{\alpha i}-\mathbf{R})
\phi_{\nu}({\bf r}-\tau_{\beta j}),
 \label{eq:denmat}
\end{eqnarray}
where $f_{n\mathbf{k}}$ is the Fermi
occupation factor, and $N_{k}$ is the number of {\bf k} points in
the Brillouin zone (BZ) sampling, which is typically equivalent to the number of
real-space unit cells $N$ in the Bloch summation.
$\rho_{\mu\nu}(\mathbf{R})$
in Eq.~\ref{eq:denmat} is the density matrix in real space, defined as
\begin{equation}
\label{eq.densitymatrix}
\rho_{\mu\nu}(\mathbf{R})
=\frac{1}{N_{k}}\sum_{n\mathbf{k}}f_{n\mathbf{k}}
c_{n\mu,\mathbf{k}}^{\ast}
c_{n\nu,\mathbf{k}}
e^{-i\mathbf{k}\cdot\mathbf{R}}.
\end{equation}
Please note that in the last line of Eq.~\ref{eq:denmat}, we have assumed, without losing
generality,  the atomic orbitals to be real, {\it i.e.}, $\phi_{\mu}^{\ast}=\phi_{\mu}$.

Given the expansion of the Kohn-Sham states in terms of atomic orbitals
in Eq.~(\ref{eq:Bloch_summation}), the Kohn-Sham equation Eq.~(\ref{eq:Kohn_Sham}) becomes a
generalized eigenvalue problem,
\begin{equation}
\label{eq.kohnsham}
H({\mathbf{k}})c_{\mathbf{k}}
=E_{\mathbf{k}} S(\mathbf{k}) c_{\mathbf{k}},
\end{equation}
where $H({\mathbf{k}})$, $S({\mathbf{k}})$ and
$c_{\mathbf{k}}$ are the Hamiltonian matrix, overlap matrix
and eigenvectors at a given ${\bf k}$ point, respectively.
$E_{\mathbf{k}}$ is a diagonal matrix whose entries are the Kohn-Sham
eigenenergies.
To obtain the Hamiltonian matrix $H({\mathbf{k}})$, we first calculate
\begin{equation}
H_{\mu\nu}({\bf R})=
\langle\phi_{\mu {\bf R}}|\hat{T}
+\hat{V}^{\rm ext}+\hat{V}^{H}+\hat{V}^{xc}|
\phi_{\nu 0}\rangle,
\label{Eq:H_realspace}
\end{equation}
where $\mu$, $\nu$ are atomic orbital indices within one unit cell, and
$\phi_{\mu {\bf R}}$=$\phi_{\mu}({\bf r}-{\bf R}-\tau_{\alpha i})$,
$\phi_{\nu 0}$=$\phi_{\nu}({\bf r}-\tau_{\beta j})$.
The Hamiltonian matrix at a given ${\bf k}$ point can be obtained as,
\begin{equation}
H_{\mu\nu}(\mathbf{k})=
\sum_{\mathbf{R}}
e^{-i\mathbf{k}\cdot\mathbf{R}} H_{\mu\nu}({\bf R})\, .
\label{eq:hk}
\end{equation}
Similarly, the overlap matrix at a given ${\bf k}$ point is obtained as,
\begin{equation}
S_{\mu\nu}(\mathbf{k})=
\sum_{\mathbf{R}}
e^{-i\mathbf{k}\cdot\mathbf{R}}S_{\mu\nu}({\bf R}),
\label{eq:sk}
\end{equation}
where,
\begin{equation}
S_{\mu\nu}(\mathbf{R})=
\langle\phi_{\mu {\bf R}}|
\phi_{\nu 0}\rangle.
\label{eq:SR}
\end{equation}
The construction of $H(\mathbf{k})$ and $S(\mathbf{k})$, as well as solving Eq.~(\ref{eq.kohnsham})
take most of the computational time. These two aspects
will be discussed in more details in Secs~\ref{sec:H} and \ref{sec:solver}.

In many cases, when the investigated unit cell is large enough,
a single $\Gamma$-point in the BZ is enough to get converged results.
In these cases, both the Hamiltonian and overlap matrices
are real symmetric matrices.
In the \CodeName~ package, we treat the $\Gamma$-point only calculations separately
to improve the efficiency.

\subsection{Systematically improvable atomic basis sets}
\label{sec:basis_set}
Before going into the construction processes of $H_{\mu\nu}({\bf R})$ and $S_{\mu\nu}({\bf R})$,
here we introduce the CGH atomic orbitals that are used in \CodeName.
The quality of atomic basis is essential to obtain accurate results.
Unlike PW basis, with which the quality of the calculations can be
systematically improved by simply increasing the PW energy cutoff,
the way to generate high-quality atomic basis functions is much more complicated.
In the last decades, considerable efforts have been devoted to developing high quality atomic orbitals.
\cite{Delley:1990,kenny00,junquera01,anglada02,ozaki03}
\CodeName~ adopts a scheme proposed by CGH\cite{chen10} to generate systematically improvable, optimized
atomic basis sets.

An atomic basis function can be written as a radial function multiplied by spherical
harmonics (in practice we use solid spherical
harmonic functions, which are real functions),
\begin{equation}
\label{eq.radial}
\phi_{lm\zeta}(\mathbf{r})=f_{l\zeta}(r)Y_{lm}(\hat{r}),
\end{equation}
where the indices $l,m$, and $\zeta$ have the usual meanings of angular momentum quantum number, magnetic quantum
number, and the multiplicity of the orbitals for $l$.
One usually needs more than one radial functions for each angular momentum to improve the quality and
transferability of the atomic basis sets.

In the CGH scheme, the radial function $f_{l\zeta}(r)$ is expanded in terms of
a set of spherical Bessel functions (SBFs), with
the coefficients of the SBFs yet to be determined, {\it i.e.},
\begin{equation}
f_{l\zeta}(\mathbf{r})=\left\{
\begin{array}{ll}
\sum_{q}c_{l\zeta q}j_l(qr), & r < r_c\\
0 & r \geq r_c \, ,\\
\end{array}
\right.
\end{equation}
where $j_{l}(qr_{c})$ is the SBF with radius cutoff $r_{c}$.
The possible $q$ values are chosen such that $j_{l}(qr_{c})$=0.
A kinetic energy cutoff is chosen to determine the maximal value
of $q$, and thus the number of SBFs.
We set the SBFs to be strictly zero beyond the radius cutoff $r_{c}$.
In fact, they have been used directly as short-ranged basis set
in first-principles calculations. \cite{haynes97,gan00}
Because (almost) any function within the radius cutoff $r_c$ can
be represented as a linear combination of SBFs,
this gives us a large number of degrees of freedom for optimizing the atomic basis set.

To obtain optimized atomic basis,
we vary the coefficients of the SBFs
to minimize the spillage between the atomic basis set
and a set of selected reference systems.
The spillage $\mathcal{S}$ is a positive number
defined as the difference between the Hilbert
space spanned the atomic basis set and
the wave functions calculated by PW basis,
\begin{equation}
\mathcal{S}=\frac{1}{N_{n}N_{\mathbf{k}}}
\sum_{n=1}^{N_{n}}\sum_{\mathbf{k}=1}^{N_{\mathbf{k}}}
\langle\Psi_{n\mathbf{k}}|1-\hat{P}_{\mathbf{k}}|\Psi_{n\mathbf{k}}\rangle,
\end{equation}
where $\hat{P}$ is a projector spanned by the atomic orbitals,
\begin{equation}
\hat{P}_{\mathbf{k}}=\sum_{\mu\nu}|\phi_{\mu,\mathbf{k}}\rangle
\mathrm{S}_{\mu\nu}^{-1}(\mathbf{k})\langle \phi_{\nu,\mathbf{k}}|.
\end{equation}
Here, $\mathrm{S}^{-1}(\mathbf{k})$
is the inverse of overlap matrix $\mathrm{S}(\mathbf{k})$
between numerical atomic orbitals.
The spillage has been proposed before to measure the quality of
a set of atomic basis. \cite{portal95,portal96,kenny00}
We then use simulated annealing method to determine the coefficients
that minimize the spillage.
Sometimes the numerical orbitals obtained from this procedure have
unphysical oscillations, which may lower the transferability
of the basis set. To eliminate these unphysical oscillations,
we further minimize the kinetic energy of each atomic orbital while keeping the
spillage almost unchanged. More details of this procedure can be found
in Ref.~\onlinecite{chen10}.

The reference systems used in the basis-generation procedure are very important to ensure
the transferability of numerical atomic orbitals.
The CGH scheme allows the users to choose freely the target systems to generate high-quality
basis sets for different purposes.
From our experience, homonulclear dimers are very good
reference systems~\cite{Delley:1990,volker09} for generating transferable
atomic orbitals for general purposes.
To avoid any possible bias of the basis set towards certain geometrical structure,
an average over dimers of several different bond lengths (compressed or elongated)
is taken as the target in the optimization procedure.
We provide scripts to generate the atomic bases using diatomic molecules as reference systems.
The scripts set up PW calculations provided in the code for the dimers at various bond lengths.
We remark that after the atomic basis sets are generated, for consistency, one must
use the same pseudopotential and energy cutoff in later atomic orbitals based calculations
as those used in the basis generation.

The CGH scheme is very flexible and easy to implement. One can choose freely
the angular momentum of the orbitals,
and the multiplicity of the radial functions for each
angular momentum. All atomic orbitals are generated from the same procedure and criteria.
These orbitals form a sequence of hierarchial basis sets, which have a systematic
convergence behavior towards the PW reference.
Furthermore, without any assumptions of the shapes of radial functions $f_{l\zeta}(r)$,
in principle we can get the fully optimized radial functions.
As will be shown in Sec.~\ref{sec:results}, the atomic orbitals generated
in this way indeed show excellent accuracy and transferability
for various systems.

\subsection{Hamiltonian and overlap matrices construction}
\label{sec:H}

As mentioned above, in order to construct $H({\bf k})$ and $S({\bf k})$ at given ${\bf k}$ points using
Eqs.~(\ref{eq:hk}) and (\ref{eq:sk}), we need to first calculate $H_{\mu\nu}({\bf R})$ and $S_{\mu\nu}({\bf R})$.
During the processes, we take the full advantage of the short-range nature of the atomic orbitals, {\it i.e.},
only the matrix elements whose corresponding atomic orbitals have non-zero overlaps are evaluated.
This is because each matrix element could be written as an integral in real-space grids,
if this integral involves two spatially well-separated basis functions that are not overlapping,
then the result of the integral should be zero.
This feature leads to a sparse matrix and then $\mathcal{O}(N)$ scaling of the number of integrals,
which is a significant advantage compared to PW based methods.

Now we briefly discuss how each term in $H_{\mu\nu}({\bf R})$ and $S_{\mu\nu}({\bf R})$ is calculated.
As is clear from Eq.~(\ref{Eq:H_realspace}), the Hamiltonian matrix has several components,
which are computed by two different techniques, namely the two-center integral technique and
the grid integral technique, respectively.
First, the kinetic energy matrix $T_{\mu\nu}({\bf R})=
\langle\phi_{\mu{\bf R}}|\hat{T}|\phi_{\nu0}\rangle$,
the non-local pseudopotential matrix $V^{NL}_{\mu\nu}({\bf R})=
\langle\phi_{\mu{\bf R}}|\hat{V}^{NL}|\phi_{\nu0}\rangle$, as well as
the overlap matrix $S_{\mu\nu}({\bf R})=
\langle\phi_{\mu{\bf R}}|\phi_{\nu0}\rangle$
can be efficiently calculated by the two-center integral technique,~\cite{sankey89}
which has been described thoroughly in Ref.~\onlinecite{soler02}.
The two-center integrals can be split into two parts:
a one-dimensional integral over radial functions and an
angular integral involving spherical harmonic functions.
The radial integrals are tabulated for a
wide range of distances between two orbitals once for all. The value
for two orbitals within a reasonable distance can then be interpolated from the table.
The angular integral involving spherical harmonic functions
leads to quantities of so-called the Gaunt coefficients, which can also be
easily calculated. Therefore, the two center integrals can be evaluated
very efficiently.\cite{soler02}
Further details on the two-center integral technique
are given in Appendix~\ref{sec:twocenter}.

The matrix elements of the local potentials,
$V^{loc}_{\mu\nu}({\bf R})=
\langle\phi_{\mu{\bf R}}|\hat{V}^{loc}|\phi_{\nu0}\rangle$,
with
\begin{equation}
V^{loc}({\bf r})=V^L({\bf r})+V^H({\bf r})+V^{xc}({\bf r})
\end{equation}
are evaluated on a uniform grid in real space.
Here $V^L({\bf r})$ is the sum of all local pseudopotentials.
We first evaluate the local potential $V^{loc}({\bf r})$ on each grid point:
the local pseudopotentials and Hartree potentials are calculated using
techniques adapted from PW basis, {\it i.e.}, they are
first calculated in the reciprocal space, and
then Fourier transformed to the real-space grid. The exchange-correlation
potential can be directly evaluated on the real-space grid. Once we have
$V^{loc}({\bf r})$, the matrix elements $V^{loc}_{\mu\nu}({\bf R})$ are directly summed over the real-space grid.
More details on the grid-based integral technique
for the local potentials are given in Appendix~\ref{sec:vlocal}.
The grid integrals are one of the most time consuming parts in the algorithms
based on atomic orbitals. However, the computation efforts of the grid integrals
only scale linearly with the system size, and can be easily parallelized.

\subsection{Kohn-Sham equation solvers}
\label{sec:solver}

After the Hamiltonian and overlap matrices are constructed,
the Kohn-Sham equations are solved separately at each {\bf k}-point, which
amounts to solving a generalized eigenvalue problem.
This constitutes the major computational bottleneck for systems
larger than a few hundreds of atoms.
Standard diagonalization method scales as $\mathcal{O}(N^{3})$,
where $N$ is the matrix dimension.
There are a few parallel matrix eigenvalue solvers available.
\CodeName~ uses a package named High Performance Symmetric Eigenproblem Solvers (HPSEPS)
developed by the Supercomputing Center of Chinese Academy of Science,
to diagonalize the Kohn-Sham Hamiltonian.\cite{hpseps}
HPSEPS provides parallel solvers for
generalized eigenvalue problems concerning large dimensions of matrix.

Recently, Lin {\it et al.}~\cite{LinLuYingE2009,LinLuYingEtAl2009,LinChenYangEtAl2013}
developed the PEXSI technique, which provides an alternative way for solving the Kohn-Sham problem
without using a diagonalization procedure. Compared to linear scaling
approach, PEXSI does not rely on the nearsightedness principle either to
truncate density matrix elements.  In a
$\Gamma$-point calculation, the
basic idea of PEXSI can be illustrated as follows. Denote by $M$ the
number of atomic orbitals,
$\Phi({\bf r})=[\phi_1({\bf r}),\cdots,\phi_{M}({\bf r})]$
the collection of all atomic orbitals in the real space, and $\hat{\gamma}({\bf r},{\bf r}')$ the single
particle density matrix in the real space. Then the PEXSI approach first expands
$\hat{\gamma}({\bf r},{\bf r}')$ using a $P$-term pole expansion as
\begin{equation}
  \begin{split}
    \hat{\gamma}({\bf r},{\bf r}') &\equiv \Phi({\bf r}) \Gamma \Phi^*({\bf r}') \\
    &\approx \Phi({\bf r}) \Im\left(
    \sum_{l=1}^{P}\frac{\omega^{\rho}_l}{H - (z_l+\epsilon_F) S}\right)
    \Phi^*({\bf r}').
  \end{split}
  \label{eqn:gammapole}
\end{equation}
Here $H,S,\Gamma$ are the Hamiltonian matrix, the overlap matrix and the
single particle density matrix represented
under the atomic orbital basis set $\Phi$, respectively. $\epsilon_F$ is
the chemical potential or Fermi energy. The complex shifts
$\{z_{l}\}$ and weights $\{\omega^{\rho}_l\}$ are
determined through a simple semi-analytic formula, and takes negligible
amount of time to compute. The number of terms of the pole expansion is
proportional to $\log(\beta\Delta E)$, where $\beta$ is the inverse of
temperature and $\Delta E$ is the spectral radius, which can be
approximated by the largest eigenvalue of the $(H,S)$ matrix pencil. The
logarithmic scaling makes the pole expansion a highly efficient approach
to expand the Fermi operator.

At first it may seem that
the entire Green's function-like object $[H - (z_l+\epsilon_F) S]^{-1}$
needs to be computed.
However, if we target the electron density
$\rho({\bf r})=\hat{\gamma}({\bf r},{\bf r})$, then only $\{\left[(H -
(z_l+\epsilon_F)
S)^{-1}\right]_{\mu\nu}\vert H_{\mu\nu}\ne 0\}$ are actually needed.
A selected inversion algorithm can be used to efficiently compute these
\textit{selected elements} of the Green's function, and therefore the
entire electron density.
The computational cost of the PEXSI technique scales at most as
$\Or(N^2$). The actual computational cost depends on the
dimensionality of the system: the cost for quasi-1D systems such as
nanotubes is $\Or(N)$ \ie{}  linear scaling; for quasi-two-dimensional systems
such as graphene and surfaces (slabs) the cost is $\Or(N^{1.5})$; for general
three-dimensional bulk systems the cost is $\Or(N^2)$.
This favorable scaling hinges on the sparse character of
the Hamiltonian and overlap matrices,
but not on any fundamental assumption about the
localization properties of the single particle density matrix.
This method is not only applicable to the
efficient computation of electron density, but also to other physical
quantities such as free energy, atomic forces, density of states and local density of
states. All these quantities can be obtained without computing any eigenvalues or eigenvectors.
For instance, the atomic force for atom $i$ in species $\alpha$ can be computed as
\begin{equation}
	F_{\alpha i}  \approx
  -\Tr\left[ \Gamma \frac{\partial H}{\partial \tau_{\alpha i}} \right]
      +\Tr\left[ \Gamma^E \frac{\partial S}{\partial \tau_{\alpha i}} \right],
  \label{eqn:forcepole}
\end{equation}
where the first part is independent of PEXSI algorithm and will be discussed
in the next subsection.
The second term in Eq.~\ref{eqn:forcepole} depends on the energy density matrix,
which is written as
\begin{equation}
  \Gamma^{E} \approx \Im \sum_{l=1}^{P}
  \frac{\omega^{E}_l}{H - (z_l+\epsilon_F) S}.
  \label{eqn:gammaF}
\end{equation}
This matrix is given again by pole expansions with \textit{the same poles}
as those used for computing the charge density, with different weights $\{\omega^{E}_l\}$.
For more detailed information we refer readers
to Refs.~\onlinecite{LinChenYangEtAl2013} and \onlinecite{LinGarciaHuhsYang2014}.
In order to use the PEXSI technique for multiple {\bf k}-point calculations,
we need to work with the Green's function of a non-Hermitian Hamiltonian
that is only structurally symmetric.
The massively parallel selected inversion method for non-Hermitian but structurally symmetric matrices are
currently under development, and will be integrated into \CodeName~ in the future
to perform large scale electron structure calculations with multiple {\bf k}-point sampling.

Compared to existing techniques, the PEXSI method has some notable features:
1) The efficiency of the PEXSI technique does not depend on the
existence of a finite Highest Occupied Molecular Orbital (HOMO)-Lowest Unoccupied Molecular Orbital (LUMO) gap,
and can be accurately applied to general materials systems including small gapped systems and metallic systems.
The method remains accurate at low temperatures.
2) The PEXSI method has a two-level parallelism structure and is by design highly scalable.
The recently developed massively parallel PEXSI technique can make efficient use of
$10,000\sim 100,000$ processors on high performance machines.  3) As a
Fermi operator expansion based method, PEXSI allows the use of a
hybrid scheme that combines density of states estimation based on Sylvester's law
of inertia with Newton's method to obtain the chemical potential. This
is a highly efficient and robust approach with
respect to the initial guess of the chemical potential,
and is independent of the presence of gap states.
4) PEXSI can be controlled with a few input parameters, and can act
nearly as a black-box substitution of the diagonalization procedure
commonly used in electronic structure calculations.

In order to benefit from the PEXSI method, the Hamiltonian and overlap
matrices must be sparse, and this requirement is satisfied when atomic
orbitals are used to discretize the Kohn-Sham Hamiltonian.  The
sequential version of PEXSI has been demonstrated before with \CodeName~, \cite{LinChenYangEtAl2013}
and the massively parallel version of
PEXSI is recently integrated with SIESTA~\cite{LinGarciaHuhsYang2014}
for studying large scale systems with more than $10,000$ atoms with
insulating and metallic characters on more than $10,000$ processors. The
parallel PEXSI method is to be integrated with \CodeName~.

\subsection{Total energy and force calculations}
\label{sec:energy_force}

Once the Kohn-Sham equation is solved, one can obtain the total energy of the system
using the Harris functional,\cite{Harris:1985}
\begin{equation}
\label{eq:totalEnergy}
E^{tot}=E^{band} - \int V^{Hxc}({\bf r})\rho({\bf r}) d{\bf r}
+E^{H}
+E^{xc}
+E^{II}\, ,
\end{equation}
where $E^{band}$ is the Kohn-Sham band energy, which is the summation
over occupied Kohn-Sham orbital energies.
$E^H$, $E^{xc}$ and
$E^{II}$ are the Hartree energy, the exchange-correlation energy, and the Coulomb
energy between ions respectively. The second term in the above equation is
the so-called double-counting energy arising from the Hartree and exchange-correlation potential,
which have been included in the band energy term.
If the Kohn-Sham equation is solved by matrix diagonalization,
then the band energy is the summation of the Kohn-Sham eigenvalues $\epsilon_{n{\bf k}}$ of all occupied bands, {\it i.e.},
\begin{equation}
E^{band}=\frac{1}{N_{k}} \sum_{n {\bf k}} f_{n{\bf k}} \epsilon_{n{\bf k}} \, .
\end{equation}
Alternatively, if the PEXSI method is chosen to be the Kohn-Sham equation solver,
the band energy is calculated as (only for $\Gamma$-point now)
\begin{equation}
E^{band}= \mathrm{Tr}[\rho H] \, ,
\end{equation}
where $\rho$ is the density matrix and $H$ is the Hamiltonian matrix.
The Hartree and exchange-correlation energies are calculated on a uniform real-space grid,
and $E^{II}$ are calculated by the Ewald summation technique. \cite{ewald21}

The forces acting on the ions are given by
the derivative of the total energy with respect to the atomic coordinates.
Analytical expressions for forces computed with atomic basis sets
are more sophisticated than those with the PW basis sets.
Besides the Feynman-Hellmann forces, Pulay forces \cite{pulay69} due to the change
of the atomic basis sets during a structural relaxation should also be considered.

Therefore, we rewrite the total energy as the sum of two parts:
$E^{tot}=E^{KS}+E^{II}$, where $E^{KS}$ is the electronic part
of the total energy in Eq.~\ref{eq:totalEnergy},
while $E^{II}$ is the energy due to the Coulomb interactions between ions.
Then the total force experienced by the $i$-ion of type $\alpha$ is
\begin{equation}
F_{\alpha i}=-\frac{\partial E^{tot}}{\partial\tau_{\alpha i}}
=-\frac{\partial E^{KS}}{\partial\tau_{\alpha i}}
-\frac{\partial E^{II}}{\partial\tau_{\alpha i}}\, .
\label{eq:totforce}
\end{equation}
After some derivations (more details are given in Appendix~\ref{sec:force}), we arrive at
\begin{equation}
\frac{\partial E^{KS}}
{ \partial \tau_{\alpha i}}
=\sum_{{\bf R}}\sum_{\mu\nu} \rho_{\mu\nu}({\bf R})
\frac{\partial H_{\mu\nu} ({\bf R}) }{\partial \tau_{\alpha i}}
+ \sum_{{\bf R}}\sum_{\mu\nu} \frac{\partial \rho_{\mu\nu}({\bf R}) }{ \partial \tau_{\alpha i}}
H_{\mu\nu}({\bf R})\, ,
\label{eq:force3}
\end{equation}
with
\begin{equation}
\label{eq:force4}
\frac{\partial H_{\mu\nu}({\bf R})}{\partial \tau_{\alpha i}}
=\langle\phi_{\mu {\bf R}} |
\frac{\partial H}{\partial\tau_{\alpha i}}|\phi_{\nu0}\rangle
+\langle \frac{\partial\phi_{\mu{\bf R}}}{\partial
\tau_{\alpha i}} |H|\phi_{\nu0}\rangle
+\langle \phi_{\mu{\bf R}} | H |\frac{\partial \phi_{\nu0} }{ \partial
\tau_{\alpha i}}\rangle\, .
\end{equation}
Note that $\mu = \{\alpha, i, l, m, \zeta \}$ is the compact index for the atomic orbitals.
The first term of the above equation is the Feynman-Hellmann force,\cite{feynman39}
whereas the rest two terms yield the so-called Pulay forces.\cite{pulay69}
Following Ref.~\onlinecite{soler02}, one can prove that the term related to $\frac{\partial \rho_{\mu\nu}({\bf R}) }{ \partial \tau_{\alpha i}}$ is
\begin{equation}
 \sum_{{\bf R}}\sum_{\mu\nu} \frac{\partial \rho_{\mu\nu}({\bf R}) }{ \partial \tau_{\alpha i}}
H_{\mu\nu} =\sum_{{\bf R}}\sum_{\mu\nu} E_{\mu\nu}({\bf R}) {\partial S_{\mu\nu}({\bf R}) \over \partial \tau_{\alpha i}}  \,,
\end{equation}
where,
\begin{equation}
E_{\mu\nu}({\bf R})
=\frac{1}{N_{k}}\sum_{n\mathbf{k}}f_{n\mathbf{k}}
\epsilon_{n\mathbf{k}}
c_{\mu n,\mathbf{k}}^{\ast}
c_{n\nu,\mathbf{k}}
e^{-i\mathbf{k}\cdot{\bf R}},
\end{equation}
is the element of ``energy density matrix''.
In the above equation,
$\epsilon_{nk}$ is the band energy for band $n$ at wave vector ${\bf k}$.
This term arises because the atomic orbitals are not orthogonal.

Similar to the total energy evaluation,
different force terms are also evaluated using different techniques to maximize the efficiency of force calculations.
Specifically, due to the long range tail of local pseudopotential in real space,
it is better to calculate it in reciprocal space using PW basis set,
and this is how Feynman-Hellmann force associated with the local pseudopotentials is implemented.
Another advantage of this implementation
is that the derivative of the local pseudopotential can be easily done in reciprocal space.
By taking advantage of the short-range character of the non-local pseudopotential operators and atomic orbitals,
the force terms including the Feynman-Hellmann force arising from the non-local pseudopotential operator,
the Pulay forces arising from the kinetic energy operator, as well as the non-orthogonal forces,
are calculated using two-center integral techniques.
Finally, the Pulay forces associated with the local potentials are evaluated by grid integrals.
Here the local pseudopotentials are first evaluated in reciprocal space and Fourier transformed
to a real-space mesh.
Technical details on the force calculations can be found in Appendix~\ref{sec:force}.

With the capability of calculating forces efficiently,
a structural relaxation can be done by searching the local minimum
in potential energy surface. Two algorithms for structural relaxations are implemented,
namely the Broyden-Fletcher-Goldfarb-Shanno (BFGS) method \cite{salomon00}
and the conjugate gradient (CG) method.\cite{hager2006}

%% file: result.tex
\section{\label{sec:results}Results}

In this section, benchmark results obtained from \CodeName~ are presented for a variety of systems,
including molecules, crystalline solids, surfaces, and defects.
The tested atomic species cover both main-group elements and transition metal elements.
In particular, the convergence of calculated physical properties are tested with respect to the size of CGH orbitals.
The tested basis sets form a hierarchy by spanning from single-$\zeta$ (SZ), double-$\zeta$ (DZ),
double-$\zeta$ plus polarization functions (DZP), to triple-$\zeta$ plus double polarization functions (TZDP),
and quadrupole-$\zeta$ plus triple polarization functions (QZTP).
In the following tests, we refer to these basis sets as atomic basis sets,
or equivalently, as linear combination of atomic orbitals (LCAO).
The references for the tested properties are chosen to be
those calculated by converged PW basis set with the same pseudopotentials.
Available experimental results are also included for comparisons.

The naming of basis sets depends on the valence electrons of each element.
For example, SZ refers to a single $s$ orbital for elements that have only $s$ valence electrons, such as
alkali metal elements. For elements that have $p$ valence electrons, SZ refers to a single $s$ orbital
plus three $p$ orbitals, such as first- and second-row non-metal elements. Furthermore, for
transition metal elements, SZ refers to one $s$ orbital, three $p$ orbitals plus five $d$ orbitals.
Here the number of orbitals on each angular momentum channel is $(2l+1)$, where $l$ is the angular
momentum quantum number. The polarization functions refer to orbitals that have higher $l$ than
the maximal one used in a SZ basis set.
Specifically, in a SZ basis set that contains only one $s$ orbital, the $p$ orbitals are referred to the polarization functions;
in a SZ basis set that has both $s$ and $p$ orbitals, the $d$ orbitals are indicated as the polarization functions.
Finally, for transition metals which use all $s$, $p$, $d$ orbitals in a SZ basis set,
the $f$ orbitals are the polarization functions.

Two more parameters are needed to define a CGH atomic orbital.
First, because CGH atomic orbitals are generated by an optimization procedure that is based on
the results from PW calculations of target systems (typically diatomic molecules here),
thus the generated set of atomic orbitals depend on a specific energy cutoff ($E_\text{cut}$)
used in PW calculations.
Second, all CGH atomic orbitals are enforced to be strictly localized within a radius $R_\text{cut}$,
beyond which the atomic orbitals are set to be exactly zero.
Table~\ref{tab:LCAO_basis} lists both parameters for 24 elements that will be
used in the followings to test physical properties of systems.

\begin{table}
\centering
\caption{\label{tab:LCAO_basis}The energy cutoff  $E_\text{cut}$ (in Ry) and radius cutoff
$R_\text{cut}$ (in Bohr) parameters of the LCAO basis functions for 24 different elements used
in this paper.}
\centering
\begin{threeparttable}
\begin{tabular}{cccccc}
\hline
\hline
Element &$E_\text{cut}$ (Ry) &$R_\text{cut}$ (Bohr) &Element &$E_\text{cut}$ (Ry) & $R_\text{cut}$ (Bohr)  \\
\cline{1-6}
H & 50 & 6 & Cl & 50 & 8 \\
Li & 30 & 12  & Ti & 100 & 10 \\
C & 50 & 8    & Fe & 100 & 10 \\
N & 50 & 8 & Cu & 100 & 8 \\
O & 50 & 8 &  Ga & 50 & 9 \\
F & 50 & 8 &  Ge & 50 & 9 \\
Na & 20 & 12 & As & 50 & 9 \\
Mg & 20 & 12 & Br & 50 & 9 \\
Al & 50 & 9 & Br & 50 & 9 \\
Si & 50 & 8 & In & 50 & 9 \\
P & 50 & 9 & Sb & 50 & 9 \\
S & 50 & 8 & I & 50 & 9 \\
\hline\hline
\end{tabular}
\end{threeparttable}
\end{table}

Compared to a PW basis set which can be systematically increased to reach arbitrary accuracy in a calculation,
the accuracy of the existing LCAO basis sets are known to be difficult to improve systematically.
However, our construction strategy for numerical atomic orbitals described in Sec.~\ref{sec:basis_set}
can in principle guarantees a systematic convergence towards the PW accuracy for the target system.
We note that, for general systems, the convergence behavior of
an atomic basis set should be checked \textit{a posteriori}.
This is carried out separately for molecules in Sec.~\ref{sec:molecule} and for solids in Sec.~\ref{sec:solids}.
Among all these tests, first of all, let us look into one numerical issue that is very common in atomic-orbital based calculations
-- the so-called eggbox effect \cite{Anglada/Soler:2006} -- which occurs when the integrals of matrix elements are evaluated on a finite,
uniformly spaced real-space grid.

\subsection{\label{sec:eggbox}Eggbox effect}

The eggbox effect refers to the artificial rippling of the ground-state total energy
as a function of the atomic displacements relative to an uniform real-space grid points.\cite{Beck:2000,Anglada/Soler:2006}
Specifically, it arises from the numerical errors in evaluating the integrals of a Hamiltonian operator
with respect to the local orbitals on a finite uniform real-space grid.
This effect is completely artificial but considerably complicates the calculations of forces acting on atoms
and phonon dispersions.
Naturally, the denser the real-space grid is -- corresponding to a higher energy cutoff in the reciprocal space,
the smaller in magnitude that the rippling of the ground-state total energy will be.
Conversely, if an atomic orbital in the real space is designed in a cautious way that
the high-energy components of its Fourier transform are suppressed,
then a less denser real-space grid is needed.
Based on this principle,  Anglada and Soler proposed an efficient filtering procedure \cite{Anglada/Soler:2006}
to effectively suppress the high-energy components of their local orbitals,
without sacrificing the locality of these basis functions.
This procedure has been shown to work well in the SIESTA package.\cite{Soler/etal:2002}
In the \CodeName~ package, our basis functions are automatically confined in the reciprocal space below a certain energy cutoff
during the construction processes.

\begin{figure}[htbp]
  \includegraphics[width=0.48\textwidth, clip]{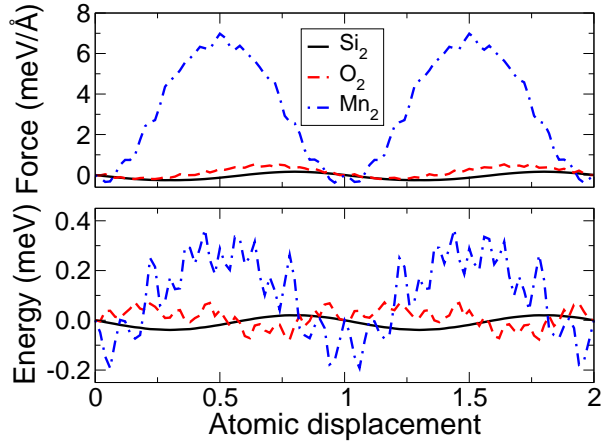}
  \caption{\label{Fig:eggbox} (Color online) The eggbox effect for Si$_2$, O$_2$, and Mn$_2$ molecules. The unit in the $x$ axis is
the spacing between neighboring real-space grid points, which amounts to 0.21 Bohr for an energy cutoff of 50 Ry.}
\end{figure}

Consequently, we show the simulated results for
diatomic molecules Si$_2$, O$_2$, and Mn$_2$ at their equilibrium distances in Fig.~\ref{Fig:eggbox}.
Unless otherwise stated, the molecular calculations are all done under the periodic boundary conditions.
A cubic box with a side length of 20 Bohr is chosen for all three molecules. The atomic basis set is chosen to be DZP.
As illustrated in Fig.~\ref{Fig:eggbox}, the eggbox effects are exceedingly small for these molecules.
In particular, for Si$_2$ and O$_2$, the oscillations of the total energy are within 1 meV, while the force oscillations are within 1 meV/\AA.
The oscillation of force for Mn$_2$ is most pronounced, but is still within 10 meV/\AA.
This accuracy is sufficient for most practical purposes.
On top of these oscillation patterns,
the additional wiggling patterns of the energy and force curves for O$_2$ and Mn$_2$
have not been well understood,
but these only occur at a smaller energy scale and have not caused any problems so far.
As will be shown in Sec.\ref{sec:surface}, accurate structural relaxations can be carried out without
further corrections for the eggbox effect.

\subsection{\label{sec:molecule}Molecules}
\begin{figure}[htbp]
  \includegraphics[width=0.48\textwidth, clip]{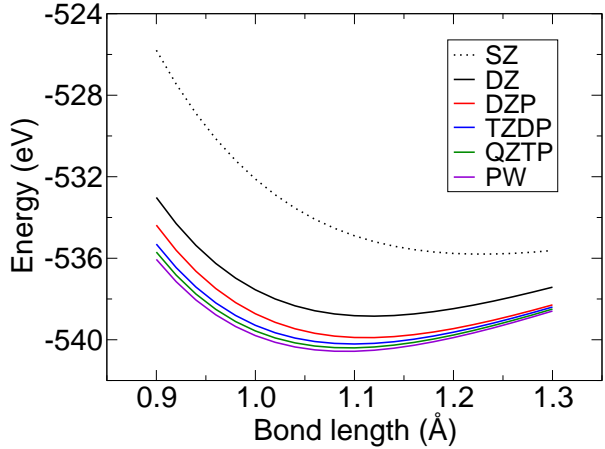}
  \caption{\label{Fig:N2} (Color online) Total energy of the N$_2$ molecule as a function of the bond length for a sequence of
  increasing LCAO basis sets. The PW results are also shown for comparison.}
\end{figure}

Having the eggbox effect under control, we now look into the convergence quality of atomic basis set
for small molecules. Similar to the eggbox test cases, a cubic cell with a side length 20 Bohr is chosen
here to avoid the artificial interactions between a molecule and its images.
Taking N$_2$ as an example, we plot its ground-state energy versus its bond distance in Fig.~\ref{Fig:N2}.
The hierarchical basis sets using in this calculation range from SZ to QZTP.
The reference is obtained from PW calculations.
As illustrated in In Fig.~\ref{Fig:N2}, as the size of the atomic basis set increases,
the total energy of the N$_2$ molecule converges systematically towards the PW limit.
From these curves, one can deduce the equilibrium bond length, atomization energy, and vibrational frequency
of the N$_2$ molecule.
These quantities obtained with atomic basis sets for a selected molecular set
can be used to validate the convergence quality of the localized basis sets,
when compared to the corresponding results obtained from PW calculations.

In the followings, we present the benchmark results for bond lengths, atomization energies,
and vibrational frequencies for 11 chemically bonded diatomic molecules.
Followed by the interaction energies of the S22 molecular test set,
\cite{Jurecka/etal:2006} obtained by the DFT-D method \cite{Grimme:2006b} as implemented in \CodeName.
The first test is used to validate our methods for chemically bonded dimers
while the second one is for weakly bonded systems.

\subsubsection{Bond lengths}
The bond length of a molecule is an important quantity.
In Table~\ref{tab:Bond_lengths}, the calculated bond lengths of 11 diatomic molecules are presented for atomic basis sets
range from SZ to QZTP. The PW results are also shown, and the experimental data are taken from Refs.~\onlinecite{Huber/Herzberg:1979} and \onlinecite{nistwebsite}.
Both LDA \cite{kohn65,Perdew/Zunger:1981} and PBE \cite{Perdew/Burke/Ernzerhof:1996} exchange-correlation
functionals are used.

As shown in Table~\ref{tab:Bond_lengths}, all the calculated equilibrium bond lengths
systematically approach the corresponding PW results with increased atomic orbital basis sets.
Specifically, the mean absolute error (MAE) for the DZP basis set is 0.018 \AA~ for both LDA and PBE calculations.
This accuracy is sufficiently good for most practical purposes.
When going beyond DZP to TZDP and QZTP basis sets, the MAEs become even smaller;
at the QZTP level, the MAE is only 0.004 \AA~ for LDA and 0.003 \AA~ for PBE.

For alkali-metal elements, the construction of high-quality localized atomic orbitals \cite{soler02,ozaki04}
is highly challenging because these orbitals tend to be very diffusive and have a longer tail
than the atomic orbitals of other elements.
However, by using CGH orbitals, we found that a rather satisfactory description of the molecular
bonding involving alkali metal atoms can be achieved, as can be seen from the examples of Na$_2$ and LiH in
Table~\ref{tab:Bond_lengths}, if a large cutoff radius, {\it i.e.,} between 10 Bohr and 12 Bohr, is used.
The same observation holds for alkali metal elements in bulk materials, as will be shown in Sec.~\ref{sec:solids}.

The experimental values in Table~\ref{tab:Bond_lengths} are only shown for comparison purpose, and not for benchmark purpose.
In all tests of bond lengths, as expected, the converged LDA bond lengths are systematically smaller than the corresponding experimental values,
while the converged PBE values show the opposite behavior.

\begin{table}
\centering
\caption{\label{tab:Bond_lengths}Bond lengths (in \AA) of diatomic molecules obtained
with various sets of atomic orbitals, in comparison with PW results and experimental data (EXP).
Both LDA and PBE results are shown.  The MAEs of atomic basis sets are obtained with reference to the PW results.}
 \centering
 \begin{threeparttable}
 \begin{tabular}{cccccccc}
 \hline
 \hline
 Molecules &SZ &DZ &DZP &TZDP &QZTP &PW &EXP \\
 \cline{1-8}
 &  \multicolumn{6}{c}{LDA} & \\[0.2ex]
 N$_2$ & 1.227 & 1.121 & 1.107 & 1.098 & 1.096 & 1.095 & 1.098 \\
 O$_2$ & 1.086 & 1.132 & 1.192 & 1.195 & 1.196 & 1.198 & 1.208 \\
 S$_2$ & 1.683 & 1.724 & 1.852 & 1.869 & 1.870 & 1.871 & 1.889 \\
 F$_2$ & 1.304 & 1.331 & 1.398 & 1.402 & 1.402 & 1.405 & 1.412 \\
 Cl$_2$ & 1.848 & 1.877 & 1.932 & 1.949 & 1.951 & 1.952 & 1.988 \\
 Br$_2$ & 2.035 & 2.184 & 2.211 & 2.226 & 2.233 & 2.240 & 2.281 \\
 I$_2$ & 2.479 & 2.563 & 2.608 & 2.623 & 2.634 & 2.641 & 2.665 \\
 Li$_2$ & 2.503 & 2.570 & 2.627 & 2.639 & 2.639 & 2.642 & 2.673 \\
 Na$_2$ & 2.901 & 2.972 & 3.028 & 3.038 & 3.041 & 3.053 & 3.079 \\
 CO & 1.271 & 1.157 & 1.136 & 1.125 & 1.125 & 1.123 & 1.128 \\
 LiH & 1.659 & 1.688 & 1.621 & 1.597 & 1.597 & 1.599 & 1.595 \\ \hline 
 MAE & 0.137 & 0.072 & 0.018 & 0.006 & 0.004 & $/$ & $/$ \\
 \hline \hline
 & \multicolumn{6}{c}{PBE} & \\[0.2ex]
 N$_2$ & 1.253 & 1.167 & 1.109 & 1.103 & 1.103 & 1.101 & 1.098 \\
 O$_2$ & 1.289 & 1.251 & 1.225 & 1.218 & 1.214 & 1.211 & 1.208 \\
 S$_2$ & 1.981 & 1.929 & 1.903 & 1.892 & 1.891 & 1.891 & 1.889 \\
 F$_2$ & 1.319 & 1.352 & 1.402 & 1.413 & 1.416 & 1.418 & 1.412  \\
 Cl$_2$ & 2.195 & 2.087 & 2.019 & 2.006 & 2.003 & 2.001 & 1.988 \\
 Br$_2$ & 2.457 & 2.396 & 2.330 & 2.313 & 2.304 & 2.292 & 2.281 \\
 I$_2$ & 2.837 & 2.782 & 2.718 & 2.693 & 2.681 & 2.674 & 2.665 \\
 Li$_2$ & 2.770 & 2.710 & 2.699 & 2.690 & 2.690 & 2.687 & 2.673 \\
 Na$_2$ & 3.265 & 3.187 & 3.103 & 3.094 & 3.094 & 3.092 & 3.079 \\
 CO & 1.140 & 1.133 & 1.130 & 1.129 & 1.129 & 1.129 & 1.128 \\
 LiH & 1.748 & 1.668 & 1.612 &1.608 & 1.607 & 1.605 & 1.595 \\ \hline 
 MAE & 0.133 & 0.065 & 0.018 & 0.007 & 0.003 & $/$ & $/$ \\
 \hline\hline
 \end{tabular}
 \end{threeparttable}
 \end{table}

\subsubsection{Atomization energies}

\begin{table}
\centering
\caption{\label{tab:Atomization_Energy}Atomization energies (in eV) of molecules obtained
with various sets of atomic orbitals, in comparison with PW and experimental results (EXP).
The MAEs of atomic basis sets are obtained with reference to the PW basis set.}
\centering
\begin{threeparttable}
\begin{tabular}{crrrrrrr}
\hline
\hline
Molecule &SZ &DZ &DZP &TZDP &QZTP &PW &EXP \\
\cline{1-8}
&  \multicolumn{6}{c}{LDA} & \\[0.2ex]
N$_2$ &    6.882 &    9.566 &   11.007 &   11.162 &   11.183 &   11.193 &    9.759 \\
O$_2$ &    5.636 &    6.495 &    7.437 &    7.491 &    7.506 &    7.542 &    5.117  \\
S$_2$ &    3.990 &    4.540 &    4.850 &    4.944 &    4.985 &    5.006 &    4.370  \\
F$_2$ &    0.746 &    0.995 &    1.834 &    1.876 &    1.886 &    1.917 &    1.601 \\
Cl$_2$ &    1.430 &    1.648 &    2.871 &    2.912 &    2.912 &    2.943 &    2.480 \\
Br$_2$ &    1.698 &    2.086 &    2.342 &    2.381 &    2.393 &    2.412 &    1.971  \\
I$_2$ &   1.179 &    1.457 &    1.931 &    1.943 &    1.965 &    1.984 &    1.542  \\
Li$_2$ &    1.323 &   1.135 &   1.111 &    1.106 &    1.104 &    1.083 &    1.037  \\
Na$_2$ &    1.134 &    0.998 &    0.964 &    0.933 &    0.933 &    0.902 &    0.735  \\
CO &   10.252 &   11.099 &   12.722 &   12.741 &   12.754 &   12.758 &   11.108  \\
LiH &    1.585 &    2.218 &    2.633 &    2.684 &    2.684 &    2.664 &    2.415  \\ \hline 
MAE & 1.408 & 0.769 & 0.080 & 0.034 & 0.022 & $/$ & $/$ \\ \hline
\hline
& \multicolumn{6}{c}{PBE} & \\[0.2ex]
N$_2$ &    6.389 &    8.727 &   10.375 &   10.592 &   10.592 &   10.623 &    9.759 \\
O$_2$ &    4.712 &    5.183 &    6.043 &    6.133 &    6.145 &    6.190 &    5.117 \\
S$_2$ &    4.032 &    4.467 &    4.602 &    4.664 &    4.747 &    4.788 &    4.370 \\
F$_2$ &    0.852 &    1.239 &    1.786 &    1.815 &    1.820 &    1.834 &    1.601 \\
Cl$_2$ &    1.552 &    2.196 &    2.682 &    2.714 &    2.721 &    2.734 &    2.480 \\
Br$_2$ &    1.329 &    1.783 &    2.024 &    2.065 &    2.086 &    2.105 &    1.971  \\
I$_2$ &   1.085 &    1.293 &    1.642 &    1.685 &    1.704 &    1.728 &    1.542  \\
Li$_2$ &   1.306  &  1.124 &   1.096  &   1.088 &    1.087 &    1.062 &    1.037 \\
Na$_2$ &   1.106 & 0.943 &    0.902 &    0.881 &    0.881 &    0.850 &    0.735  \\
CO &    10.622 &   10.969 &   11.518 &   11.569 &   11.576 &   11.609 &   11.108  \\
LiH &    2.975 &    2.684 &    2.643 &    2.622 &    2.612 &    2.601 &    2.415  \\ 
\hline
MAE & 1.083 & 0.545 & 0.097 & 0.041 & 0.026 & $/$ & $/$ \\
\hline\hline
\end{tabular}
\end{threeparttable}
\end{table}
The atomization energy refers to the energy cost to split a molecule into individual atoms.
It is an important property in thermochemistry.
The benchmark for the atomization energy are done using the same diatomic molecular
set and hierarchal atomic orbitals as used for bond-length tests.
The results are presented in Table~\ref{tab:Atomization_Energy}.
For open-shell atoms, the spin-polarized, symmetry-broken solution usually has the lowest ground-state
energy, and is thus taken here as the reference of calculated atomization energy.
As can be seen from Table~\ref{tab:Atomization_Energy}, for all molecules,
the atomization energies obtained with the atomic basis sets converge systematically towards the PW limit.

Quantitatively, however, the accuracy of atomization energies obtained with atomic basis sets
is not so spectacular as the one of geometrical properties. In particular,
the atomization energies from SZ basis are indeed too small;
the corresponding MAE is over 1 eV.
The MAE is improved by about a factor of 2 when the basis set goes from SZ to DZ, but is still far from satisfactory.
A dramatic improvement is achieved at the DZP level where a MAE around $0.1$ eV is an acceptable accuracy for most practical purposes.
Nevertheless, to reach the ``chemical accuracy" (1 kcal/mol $\approx$ 0.043 eV), one has to go further in the hierarchy of atomic basis sets.
The ``chemical accuracy" is almost reached with the TZDP basis (0.034 eV for LDA and 0.041 eV for PBE)
and well reached with QZTP (0.022 eV for LDA and 0.026 eV for PBE).
Here, the so-called basis-set error is much smaller than the errors of the energy functionals.
For this diatomic molecular test set listed in Table~\ref{tab:Atomization_Energy},
on average PBE overbinds by 0.36 eV and LDA overbinds by 0.75 eV
by comparing the converged PW results to the experimental ones.

\subsubsection{Vibrational frequency}
For a given molecule, the equilibrium bond length and the atomization energy reflect the position
and depth of the minimum in its potential energy surface, respectively,
and the vibrational frequency probes the curvature
of the potential energy surface around the equilibrium geometry.
The vibrational frequency can be measured
directly by experiment and is a powerful probe of structural and bonding
characteristics of a molecule.

In Table~\ref{Vibrational_Frequency}, we present the calculated vibrational frequencies
for the same molecular test set and basis sets used before.
Compared to the PW reference results, the MAE is reduced dramatically when the basis set goes from SZ to DZP.
Specifically, at the DZ level the MAEs are 65 cm$^{-1}$ for LDA and 47 cm$^{-1}$ for PBE,
and these numbers are further reduced to 6 cm$^{-1}$ when the basis set increased from the DZ to DZP.
The results indicate the importance of polarization functions in an atomic basis set to
accurately describe the curvature of the potential energy surface.
Nevertheless, unlike the tests for bond lengths and atomization energies of molecules,
increasing the size of basis set does not guarantee a better description of these vibrational frequencies.
We thus conclude that when CGH basis set goes beyond DZ, it still introduces around 6 cm$^{-1}$ error
for the vibrational frequencies of molecules comparing to PW calculation results.
Overall speaking, this accuracy is already excellent for most practical purposes.

\begin{table}
\centering
\caption{\label{Vibrational_Frequency}Vibrational frequencies (in cm$^{-1})$ of molecules obtained
with various sets of atomic orbitals, in comparison with PW and experimental results (EXP).
Results from both LDA and PBE are shown.
The MAEs are calculated based on the results from atomic basis set and PW basis set.}
\centering
\begin{threeparttable}
\begin{tabular}{cccccccc}
\hline
\hline
Molecules &SZ &DZ &DZP &TZDP &QZTP &PW &EXP \\
\cline{1-8}
&  \multicolumn{6}{c}{LDA} & \\[0.2ex]
N$_2$ & 1857 & 2261 & 2378 & 2394 & 2391 & 2370 & 2359 \\
O$_2$ & 1246 & 1329 & 1558 & 1581 & 1582 & 1567 & 1580 \\
S$_2$ & 517 & 675 & 717 & 719 & 719 & 721 & 726 \\
F$_2$ & 813 & 885 & 920 & 915 & 917 & 924 & 917 \\
Cl$_2$ & 498 & 524 & 561 & 560 & 562 & 567 & 560 \\
B2$_2$ & 471 & 382 & 341 & 341 & 339 & 332 & 325 \\
I$_2$ & 345 & 234 & 199 & 199 & 201 & 206 & 215 \\
Li$_2$ & 306 & 331 & 342 & 340 &  340 & 339 & 351 \\
Na$_2$ & 129 & 141 & 151 & 148 & 150 & 147 & 159 \\
CO & 1631 & 1998 & 2130 & 2130 & 2130 & 2130 & 2170 \\
LiH & 1107 & 1286 & 1327 & 1330 & 1334 & 1341 & 1406 \\ 
\hline
MAE & 207 & 69 & 6 & 8 & 7 & $/$ & $/$ \\
\hline \hline
& \multicolumn{6}{c}{PBE} & \\[0.2ex]
N$_2$ & 1938 & 2397 & 2374 & 2378 & 2380 & 2365 & 2359 \\
O$_2$ & 1297 & 1414 & 1567 & 1567 & 1566 & 1572 & 1580 \\
S$_2$ & 546 & 684 & 721 & 718 & 716 & 723 & 726 \\
F$_2$ & 758 & 872 & 914 & 917 & 918 & 920 & 917 \\
Cl$_2$ & 427 & 517 & 562 & 566 & 564 & 563 & 560 \\
B2$_2$ & 459 & 374 & 339 & 340 & 333 & 329 & 325 \\
I$_2$ & 318 & 249 & 224 & 224 & 222 & 219 & 215 \\
Li$_2$ & 313 & 335 & 345 & 345 & 344 &  342 & 351 \\
Na$_2$ & 132 & 145 & 151 & 151 & 150 & 149 & 159 \\
CO & 1728 & 2075 & 2147 & 2149 & 2147 & 2144 & 2170 \\
LiH & 1104 & 1296 & 1332 & 1339 & 1341 & 1348 & 1406 \\ 
\hline
MAE & 192 & 48 & 6 & 6 & 5 & $/$ & $/$ \\
\hline\hline
\end{tabular}
\end{threeparttable}
\end{table}

\subsubsection{Weak interaction energy}

In the previous tests, we have demonstrated the convergence behavior of our basis sets for
chemically bonded diatomic molecules. In a sense, the good performance of these atomic basis sets
for the tested molecules is not surprising,
since the homo-nuclear diatomic molecules are the target systems when generating atomic orbitals.
Now we turn to the study of bigger and more weakly bonded molecules -- the S22 test set. \cite{Jurecka/etal:2006}
This test set contains 22 weakly interacting molecular complexes that include hydrogen
bonding, dispersion interaction, and mixed bonding types.
Since its inception, the S22 test set has been widely used to benchmark computational methods that deal
with van der Waals (vdW) interactions.

Each member in the S22 test set is a molecular dimer that contains two monomers
interacting with each other. The interaction energy is defined as the
difference between the total energy of the dimer and the sum of the total energies
of the two individual monomers in their fully relaxed geometries.
In this work, the interaction energies of the S22 molecules are
calculated using the DFT-D2 method of Grimme,\cite{Grimme:2006} specifically
PBE-D2, as recently implemented in \CodeName.

\begin{figure}[t]
\includegraphics[width=0.48\textwidth, clip=true]{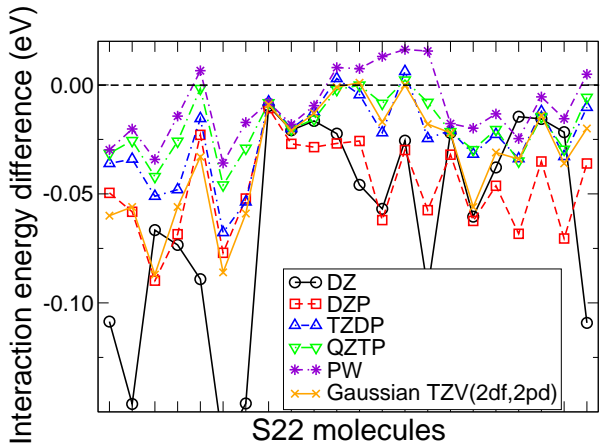}
\caption{(Color online) The PBE-D2 interaction energies of the S22 molecules obtained with
increasing LCAO basis sets, with reference to the CCSD(T) results (the zero dash line).
The results from PW basis set and Gaussian TZV(2df,2pd) (Ref.~\onlinecite{Grimme:2006}) basis set
are shown for comparison. Lines are used to guide the eye.}
\label{S22}
\end{figure}

The calculations for S22 molecules are still done with the supercell approach with cubic boxes that are
sufficiently large (up to 50 Bohr side length) to avoid artificial interactions between
the molecule and it the periodic images.  Fig.~\ref{S22} presents the interaction energy
differences of the S22 molecules calculated from two methods. The first is PBE-D2 with various atomic basis sets
and PW basis set, while the second is the coupled-cluster theory with singles, doubles,
and perturbative triples [CCSD(T)]\cite{Takatani/etal:2010} in the complete basis set limit.
The results of Grimme as reported in Ref.~\onlinecite{Grimme:2006c}, obtained using the
Gaussian TZV(2df,2pd) basis set,\cite{Schafer/Huber/Ahlrichs:1994} are also plotted in Fig.~\ref{S22} for comparison.
The CCSD(T) reference is indicated in Fig.~\ref{S22} by the dash line at energy zero.
It can be seen that by increasing the number of atomic orbitals,
the results from atomic orbitals systematically approach the PW results,
and on average get closer to the CCSD(T) references.
This trend again validates the transferability of our atomic basis set.
The quality of the Gaussian TZV(2df,2pd) is somewhere between the qualities of DZP and TZDP basis sets.
Regarding the performance of the PBE-D2 method itself,
it can be concluded that the \CodeName~ package with atomic basis set is very suitable for describing vdW forces.

\subsection{\label{sec:solids}Solids}
In Sec.~\ref{sec:molecule}, we have validated the accuracy of \CodeName~ with its associated
atomic basis sets for molecular properties.
Here we turn to the test of crystalline solids.
This is a crucial check for the transferability of the atomic basis sets,
because they are generated from diatomic systems and are now used to test solid systems.
In analogy to the bond lengths, atomization energies, and vibrational frequencies for molecules,
here we benchmark the lattice constants, cohesive energies, and bulk moduli for solids.

Twenty crystalline solids are chosen as a test set that covers group III-V and group IV
semiconductors, alkaline and alkaline-earth metals, alkaline chloride, as well as transition metals.
The lattice constants and bulk moduli of group III-IV and group IV semiconductors
obtained using \CodeName~ have already been reported in Ref.~\onlinecite{chen10}.
They are included here for completeness.
Since here we are mainly interested in the convergence behavior of the LCAO basis sets
instead of the performance of the exchange-correlation functionals, in this test only
the LDA is used for simplicity, except for Fe we choose PBE which yields
the correct body-centered-cubic (bcc) ground-state crystal structure.
For transition metal elements, TZ ($3s3p3d$) and QZ $(4s4p4d$) basis sets are used
because adding the polarization functions ($f$ orbitals) does not make a noticeable difference.
Note that these TZ/QZ basis sets are grouped together with other TZDP/QZTP basis sets
for the statistical error analysis of energy differences in Table~\ref{tab:Lattice_constant}-\ref{Bulk_Moduli}.
Similar to the molecular test case, PW basis results are also reported in these tables as references
using the same energy cutoffs and pseudopotentials as in LCAO calculation.
The MAEs are obtained between the results from atomic basis sets and the PW basis set.
Both simulations were carried by \CodeName.

Monkhorst-Pack (MP) {\bf k}-point meshes are used for the BZ sampling. Specifically,
a $4\times 4 \times 4$  {\bf k}-point mesh is used for semiconductors range from GaAs to Ge.
A $6\times 6 \times 6$ {\bf k}-point mesh is used for LiH, NaCl, and  MgO.
A denser $10\times 10 \times 10$ {\bf k}-point mesh is used for metals include bcc Na, face-centered-cubic (fcc) Al, fcc Cu, and bcc Fe.
Finally,  a $10\times 10 \times 7$ {\bf k}-point mesh is used for hexagonal-close-packed (hcp) Ti.
For transition metals, the semi-core electrons are treated as valence electrons for Ti, but not for Fe and Cu.
For the geometry optimization of Ti hcp structure,
the optimal c/a ratio is 1.590 by determining the minimum of a two dimensional (a, c) energy landscape,
with $a$ and $c$ being the lengths of the lattice vectors in the hcp structure.

\subsubsection{Lattice constants}
Table~\ref{tab:Lattice_constant} shows the lattice constants of tested crystalline solids
obtained with our hierarchical LCAO basis sets,
in comparison with the PW results, as well as experimental data
as collected in Refs.~\onlinecite{Harl/Schimka/Kresse:2010,Schimka/etal:2013}.
As illustrated in Table~\ref{tab:Lattice_constant}, the calculated lattice constants of solids
converge systematically with respect to the LCAO basis set size.
For the tested solids, an MAE of 0.02 \AA~ can be achieved with the DZP basis set,
while an MAE of 0.01 \AA~ can be reached at the level of TZ(DP) basis set.
This accuracy is comparable to that achieved for bond lengths of molecules,
and confirms that the structural properties are well described with the LCAO basis sets.
In addition, it can be seen that the convergence behavior of the basis sets for simple and transition metal elements
are very similar to that for group III-V and group IV elements.

\begin{table}[htbp]
\centering
\caption{\label{tab:Lattice_constant}Lattice constants (in \AA) of 20 solids obtained from
various LCAO basis sets, compared to the PW and experimental (EXP) results.
Experimental data (corrected for zero-point anharmonic effects) are taken from
Refs.~\onlinecite{Harl/Schimka/Kresse:2010} and \onlinecite{Schimka/etal:2013}.}
\centering
\begin{threeparttable}
\begin{tabular}{lcccccccc}
\hline
\hline
Solid &SZ &DZ &DZP &TZ(DP) &QZ(TP) &PW &EXP \\
\hline
GaAs & 5.63 & 5.59 & 5.57 & 5.55 & 5.55 & 5.54 & 5.64 \\
GaP & 5.43 & 5.40 & 5.35 & 5.33 & 5.34 & 5.34 & 5.44 \\
GaN & 4.30 & 4.35 & 4.40 & 4.41 & 4.41 & 4.41 & 4.52 \\
InAs & 6.01 & 5.97 & 5.97 & 5.96 & 5.96 & 5.96 & 6.05 \\
InP & 5.84 & 5.81 & 5.79 & 5.78 & 5.78 & 5.78 & 5.86 \\
InSb & 6.46 & 6.40 & 6.39 & 6.39 & 6.39 & 6.38 & 6.47 \\
AlAs & 5.76 & 5.70 & 5.62 & 5.61 & 5.61 & 5.60 & 5.65 \\
AlP & 5.55 & 5.51 & 5.42 & 5.41 & 5.41 & 5.40 & 5.45 \\
AlN & 4.39 & 4.33 & 4.29 & 4.27 & 4.27 & 4.27 & 4.37 \\
C & 3.63 & 3.55 & 3.51 & 3.50 & 3.50 & 3.50 & 3.55 \\
Si & 5.59 & 5.53 & 5.41 & 5.40 & 5.40 & 5.40 & 5.42 \\
Ge & 5.73 & 5.69 & 5.64 & 5.61 & 5.61 & 5.61 & 5.64 \\
LiF & 4.15 & 3.94 & 3.90 & 3.88 & 3.88 & 3.88 & 3.97 \\
NaCl & 5.62 & 5.54 & 5.51 & 5.50 & 5.50 & 5.50 & 5.57 \\
MgO & 3.96 & 4.02 & 4.06 & 4.07 & 4.07 & 4.07 & 4.19 \\
Na (bcc) & 3.51 & 3.94 & 4.05 & 4.06 & 4.06 & 4.08 & 4.21 \\
Al (fcc) & 3.71 & 3.81 & 3.87 & 3.90 & 3.91 & 3.93 & 4.02 \\
Cu (fcc) & 3.26 & 3.51 & $/$  & 3.52 & 3.53 & 3.53 & 3.60 \\
Fe (bcc) & 2.45 & 2.58 & $/$  & 2.72 & 2.72 & 2.73 & 2.86 \\
Ti (hcp) & 2.61 & 2.78 & $/$  & 2.80 & 2.80 & 2.81 & 2.96 \\ \hline 
MAE & 0.16 & 0.07 & 0.02 & 0.01 & 0.01 & $/$ & $/$ \\ \hline\hline
\end{tabular}
\end{threeparttable}
\end{table}

\subsubsection{Cohesive energies}
\begin{table}[htbp]
\centering
\caption{\label{Cohesive_energies} Cohesive energies (in eV/atom) of 20 solids at their equilibrium
lattice constants obtained from various LCAO basis sets, compared to the PW and experimental (EXP) results.
The experimental data are taken from Ref.~\onlinecite{NISTsolids},
corrected for the zero-temperature vibration effect as done in Ref.~\onlinecite{Harl/Schimka/Kresse:2010}.}
\centering
\begin{threeparttable}
\begin{tabular}{ccccccccc}
\hline
\hline
Solid &SZ &DZ &DZP &TZ(DP) &QZ(TP) &PW &EXP \\
\hline
GaAs & 3.14 & 3.73 & 3.99 & 4.01 & 4.01 & 4.01 & 3.34 \\
GaP & 3.42 & 4.01 & 4.16 & 4.18 & 4.18 & 4.18 & 3.61 \\
GaN & 4.33 & 5.03 & 5.25 & 5.27 & 5.28 & 5.28 & 4.55 \\
InAs & 2.72 & 3.47 & 3.78 & 3.82 & 3.83 & 3.84 & 3.08 \\
InP & 3.15 & 3.95 & 4.17 & 4.21 & 4.21 & 4.22 & 3.47 \\
InSb & 2.47 & 3.20 & 3.48 & 3.53 & 3.55 & 3.55 & 2.81 \\
AlAs & 3.69 & 4.18 & 4.34 & 4.57 & 4.37 & 4.37 & 3.82 \\
AlP & 4.08 & 4.57 & 4.75 & 4.77 & 4.77 & 4.77 & 4.32 \\
AlN & 5.34 & 6.29 & 6.46 & 6.48 & 6.49 & 6.49 & 5.85 \\
C & 8.06 & 8.53 & 8.72 & 8.74 & 8.75 & 8.75  & 7.55 \\
Si & 4.08 & 4.76 & 5.20 & 5.22 & 5.24 & 5.25 & 4.68 \\
Ge & 3.75 & 4.24 & 4.55 & 4.58 & 4.59 & 4.60  & 3.92 \\
LiF & 4.17 & 4.62 & 4.88 & 4.90 & 4.90 & 4.90 & 4.46 \\
NaCl & 2.49 & 3.20 & 3.51 & 3.52 & 3.52 & 3.56 & 3.34 \\
MgO  & 4.63 & 5.46 & 5.72 & 5.74 & 5.74 & 5.77 & 5.20 \\
Na(bcc) & 0.98 & 1.10 & 1.25 & 1.27 & 1.27 & 1.27 & 1.12 \\
Al(fcc) & 3.18 & 3.79 & 3.95 & 3.96 & 3.96 & 3.97 & 3.43 \\
Cu(fcc) & 3.65 & 4.17 & $/$ & 4.46 & 4.48 & 4.50 & 3.52 \\
Fe(bcc) & 5.54 & 5.18 & $/$ & 5.10 & 5.08 & 5.08 & 4.30  \\
Ti(hcp) & 4.82 & 5.24 & $/$ & 5.33 & 5.35 & 5.36 & 4.88 \\ \hline
MAE & 0.85 & 0.26 & 0.04 & 0.01 & 0.01 & $/$& $/$ \\ \hline\hline
\end{tabular}
\end{threeparttable}
\end{table}

The cohesive energies of the tested solids at their equilibrium lattice constants are presented in Table~\ref{Cohesive_energies}.
Again, the general trend that the LCAO results systematically approach the PW limit can be observed.
It is also interesting to point out that the LCAO basis sets
show even better convergence behavior for cohesive energies of solids than atomization energies of molecules.
For example, at the DZP level, the MAE is only 0.04 eV for the cohesive energies of solids,
compared to 0.09 eV for the atomization energies of molecules.
The same behavior holds for higher levels of basis sets such as TZDP and QZTP,
where the MAEs of 0.01 eV for solids are also twice smaller than those of molecules.

\subsubsection{Bulk Moduli}
The bulk modulus is another key property of solids, reflecting the variation of the ground-state energy
with respect to the unit cell volume around the equilibrium state.
In Table~\ref{Bulk_Moduli}, the bulk moduli of the tested solid are presented.
Compared to the PW results, the SZ basis set yields a large MAE of 21.2 GPa.
However, this is quickly reduced to 5.8 GPa at DZ level and 2.1 GPa at DZP levels,
this accuracy is sufficiently accurate for most purposes.
For basis sets larger than the DZP basis set,
the MAEs are further reduced to around 1.0 GPa, which are highly accurate.
This convergence behavior of the LCAO basis sets is similar to what was observed for the
calculated vibrational frequencies for tested molecules in Sec.~\ref{sec:molecule}.
In addition, similar to the cases of testing lattice constants and cohesive energies,
the quality of the basis sets is equally good for transition metal elements
as for main-group elements.
\begin{table}[htbp]
\centering
\caption{\label{Bulk_Moduli}Bulk Moduli (in GPa) of 20 solids at their equilibrium
lattice constants obtained from various LCAO basis sets, compared to the PW and experimental (EXP) results.
The experimental data are taken from Refs.~\onlinecite{Harl/Schimka/Kresse:2010} except for Fe and Ti
[Ref.~\onlinecite{CRChandbook}]. }
\centering
\begin{threeparttable}
\begin{tabular}{cccccccc}
\hline
\hline
Solid &SZ &DZ &DZP &TZ(DP) &QZ(TP) &PW &EXP \\
 \hline
GaAs & 66.5 & 68.3 & 75.1 & 76.8 & 76.8 & 77.2 & 76 \\
GaP & 80.7 & 82.3 & 89.8 & 94.1 & 93.7 & 92.4 & 89 \\
GaN & 187.5 & 203.8 & 209.6 & 207.3 & 207.9 & 208.2 & 210\\
InAs & 59.5 & 65.8 & 67.2 & 67.3 & 67.2 & 67.9 & 60 \\
InP & 74.3 & 75.8 & 77.2 & 79.5 & 80.3 & 81.4 & 71\\
InSb & 45.0 & 48.6 & 49.9 & 49.9 & 50.2 & 51.4 & 47\\
AlAs & 62.9 & 71.4 & 74.2 & 74.6 & 74.6 & 75.3 & 77 \\
AlP & 70.1 & 78.5 & 86.2 & 87.4 & 87.7 & 88.3 & 86\\
AlN & 192.9 & 202.7 & 205.3 & 208.4 & 207.8 & 207.2 & 202\\
C & 428.7 & 442.8 & 459.3 & 459.4 & 458.5 & 458.2 & 443 \\
Si & 72.3 & 78.2 & 93.9 & 93.9 & 93.5 & 93.3 & 99  \\
Ge & 53.4 & 67.3 & 71.1 & 72.0 & 72.8 & 73.5 & 76 \\
LiF  & 44.1 & 58.5 & 60.2 & 61.1 & 61.3 & 62.6 & 70\\
NaCl  & 14.7 & 22.1 & 29.4 & 29.6 & 29.8 & 30.1 &27 \\
MgO  & 128.6 & 165.4 & 167.7 & 171.9 & 171.3 & 170.2 & 165\\
Na(bcc)  & 5.9 & 6.8 & 7.0 & 7.0 & 7.0 & 7.1 & 8 \\
Al(fcc)  & 70.2 & 74.9 & 75.8 & 76.0 & 76.0 & 76.2 & 79 \\
Cu(fcc)  & 120.8 & 147.3 & $/$ & 149.4 & 150.2 & 150.6 & 142 \\
Fe(bcc)  & 96.8 & 160.4 & $/$ & 161.1 & 161.8 & 165.0 & 170 \\
Ti(hcp)  & 52.8 & 113.3 & $/$  & 122.4 & 122.9 & 124.5 & 110 \\ \hline 
MAE   & 21.2 & 5.8 & 2.1 & 1.2 & 0.8 & $/$ & $/$ \\
\hline\hline
\end{tabular}
\end{threeparttable}
\end{table}

\subsection{\label{sec:surface} Si(100) surface reconstruction}
In the previous tests for molecules and solids,
we have established the reliability of the \CodeName~ package and the accuracy of its associated LCAO basis sets.
Here we test a more ``challenging" problem -- the reconstruction of the Si(100) surface.
This surface is technologically important for fabricating silicon-based devices
and has been intensively studied both theoretically and experimentally.
Different reconstruction models for the Si(100) surface have been proposed in the past,
\cite{Levine:1973,Appelbaum/etal:1977,Chadi:1979,Inoue/etal:1994}
and the energy hierarchy among these different reconstructions have been examined by both DFT (LDA) calculations \cite{Ramstad/Brocks/Kelly:1995}
and by quantum Monte Carlo \cite{Healy/etal:1995} calculations.
The small magnitude of energy differences between these reconstructions offers an excellent testing ground
for \CodeName~ with the LCAO basis sets.

\begin{figure}[t]
\includegraphics[width=0.48\textwidth, clip]{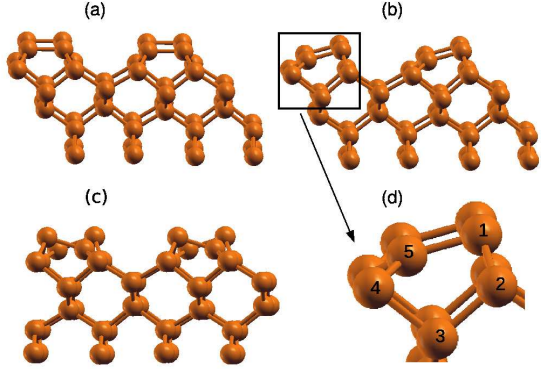}
\caption{\label{Si100_recons}(Color online) Side view of three reconstruction structures for the Si(100) surface:
(a) p($2 \times 1$)s; (b) p($2 \times 1$)a; (c) p($2 \times 2$);
(d) zoom-in of the top three layers of the p($2 \times 1$)a reconstruction structure.}
\end{figure}

Three reconstructions of the Si(100) surface are considered in the work,
namely the p$(2\times 1)$ symmetric [denoted as
p$(2\times 1)$s below], p$(2\times 1)$ asymmetric [p$(2\times 1)$a], and p$(2\times 2)$ reconstructions,
as shown in Fig.~\ref{Si100_recons}(a-c). It is well known that the Si atoms on the surface layer form dimers to lower the
energy of the system by removing one of the two dangling bonds. \cite{Schlier/Farnsworth:1959}
In the p$(2\times 1)$s reconstruction structure,
the two atoms on the top layer come closer symmetrically, with bond length between them becomes
slightly shorter than the nearest neighbor distance in the Si bulk.
In the p$(2\times 1)$a case, the dimers buckle out of the Si(100) surface.
Finally, in the p$(2\times 2)$ case, the buckled dimers change their orientations alternatively (see Fig.~\ref{Si100_recons}(c)).
Note that another c$(2\times 4)$ reconstruction exists where the adjacent buckled dimer rows orientate oppositely.
However, the energy lowering of this reconstruction is almost identical to that of p$(2\times 2)$ when using DFT with LDA, \cite{Ramstad/Brocks/Kelly:1995}
thus this fourth reconstruction structure is not considered in this work.

Next we describe the computational setup of our simulations.
\CodeName~ is used with both LCAO DZP basis set and PW basis set.
Also, the Quantum ESPRESSO (QE) package \cite{qe2009} is used, which serves as an independent check for the \CodeName~ package.
The same pseudopotential for Si is used for all three simulations.
We model the Si(100) surface by using a repeated slab which contains 12 atomic layers,
because the structural distortion below the surface layer extends 4-5 layers into the bulk as noticed before. \cite{Appelbaum/Hamann:1978}
The central two layers are fixed during the structural relaxation, and
only atoms in the outermost five layers on each side are allowed to relax.
The conjugate gradient algorithm is used for structural relaxation with a force threshold of 0.01 eV$/$\AA.
The slabs are separated by a vacuum of 30 \AA~thick to avoid artificial interactions between neighboring slabs.
When computing the energy differences of the p$(2\times 1)$s and p$(2\times 1)$a reconstructions with respect to the ideal surface,
the (2$\times$1) unit cell in the x-y plane is used, while in the case of the p$(2\times 2)$ reconstruction,
the (2$\times$2) unit cell is used for both relaxed and ideal surfaces.
In all calculations,  a $(6\times 6 \times 1)$ {\bf k}-point mesh is used.

Table~\ref{Si_surface_energy} reports the energy lowings of the p$(2\times 1)$s, p$(2\times 1)$a, and p$(2\times 2)$
reconstructions with respect to the ideal Si(100) surface.
For the ideal Si(100) surface, we fix the surface structure using the bond lengths in Si bulk.
Table~\ref{Si_surface_energy} shows an excellent agreement between the \CodeName/PW results and the QE results.
First, from QE calculations, the successive energy decreases of the three relaxation steps, {\it i.e.,}
from the ideal surface to the p(2 $\times$ 1)s, p(2 $\times$ 1)a, and p(2 $\times$ 2) reconstructed surfaces,
are 1.504 eV, 0.086 eV, and 0.063 eV per Si dimer, respectively.
These numbers can be compared to the 1.80 eV, 0.12 eV, and 0.05 eV ones as reported in Ref.~\onlinecite{Ramstad/Brocks/Kelly:1995}.
The differences between results from QE and Ref.~\onlinecite{Ramstad/Brocks/Kelly:1995}
are presumably due to the usage of different pseudopotentials, energy cutoff and BZ ${\bf k}$-point sampling.
However, both methods lead to the same energy orderings of the three reconstructions with respect to the ideal Si(100) surface.
Also it is assuring that \CodeName~ with the PW basis set gives almost identical results compared to QE calculations as can be seen in Table~\ref{Si_surface_energy}.
Finally, \CodeName~ with DZP basis set yields 1.481 eV, 0.078 eV, 0.070 eV per Si dimer energy lowerings,
which are in excellent agreement with the other two PW results.

\begin{table}[htbp]
\centering
\caption{\label{Si_surface_energy}The energy lowerings per dimer (in eV) of three different reconstructions
of the Si(100) surface, with respect to the ideal Si(100) surface.}
\centering
\begin{threeparttable}
\begin{tabular}{lccc}
\hline\hline
 Surface &  \CodeName/DZP   &  \CodeName/PW & Quantum ESPRESSO \\
\hline
p$(2\times 1)$s  & -1.481  &  -1.505 & -1.504   \\
p$(2\times 1)$a  & -1.559  &  -1.591 & -1.590  \\
p$(2\times 2)$   & -1.629  & -1.652 & -1.653  \\
\hline\hline
\end{tabular}
\end{threeparttable}
\end{table}

Next we examine the relaxed structures obtained by the three methods.
We choose the p($2\times 1$) reconstruction because its
structure distortion is most pronounced among the three.
Specifically, we look at the pentagon pattern formed by the atoms from the top three layers of
the p($2\times 1$) structure, as illustrated in Fig.~\ref{Si100_recons}(d).
In Table~\ref{tab:Si_pentagon_struc}, we listed the bond lengths and bond angles (the
five edge lengths and angles) of the pentagon as yielded by the three types of calculations.
Again, an almost perfect agreement is obtained between the results obtained by the \CodeName/PW
calculations and the QE calculation.
In this case, the bond lengths from the two methods agree within 0.001 \AA,
while the bond angles agree within $0.1$ degree (Deg).
The results from the DZP basis set are also in excellent agreement with the PW results.
The bond lengths from former calculation are slightly longer than the latter ones by 0.003-0.007 \AA,
and the bond angles differ by 0.4 Deg at most.
In conclusion, for all the tested properties of the three reconstructions of Si(100) surface,
the \CodeName~ package with the DZP basis set gives accurate results compared to PW results,
demonstrating again the ability of \CodeName~ to do reliable surface calculations.

\begin{table}[htbp]
\centering
\caption{\label{tab:Si_pentagon_struc}The bond lengths between the atoms in the top three layers of the
p$(2\times 1)$a reconstruction. $a_{ij}$ is the bond length (in \AA) between the $i$-th and $j$-th
atoms as shown in Fig.~\ref{Si100_recons}(d).
$\theta_i$ is the bond angle (in Deg.) formed by the two bonds sharing the $i$-th atom.  }
\centering
\begin{threeparttable}
\begin{tabular}{lccc}
\hline\hline
Parameter  &  \CodeName/DZP   &  \CodeName/PW & Quantum ESPRESSO \\
\hline
a$_{12}$  & 2.366  &2.360  & 2.360   \\
a$_{23}$  & 2.337 &2.334  &  2.335  \\
a$_{34}$  & 2.405 & 2.401 & 2.401    \\
a$_{45}$  & 2.307 & 2.303 & 2.302  \\
a$_{51}$  & 2.290 & 2.283 & 2.283   \\
$\theta_1$ & 90.13    & 89.88     & 89.94    \\
$\theta_2$ & 104.55   & 104.24     & 104.21    \\
$\theta_3$ & 100.19   & 99.51     & 100.49    \\
$\theta_4$ & 81.34    & 80.42     & 80.48    \\
$\theta_5$ & 121.80   & 122.15     & 122.09    \\
\hline\hline
\end{tabular}
\end{threeparttable}
\end{table}

\subsection{N defect in bulk GaAs}
Real materials contain various types of defects, and their presence
greatly affects, and often decisively determines the physical properties of materials.
In recent years, DFT-based first-principles approaches have emerged as powerful tools for
describing and understanding of point defects in solids, \cite{Freysoldt/etal:2014}
and is becoming an indispensable complement to experiments that are often
difficult and expensive to carry out.
Since usually a large supercell is required to model defects in solids,
the LCAO technique, which scales favorably with the system size,
is a preferable choice for simulating the electronic structures of defects.

As a specific example, we employ the \CodeName~ package to study the group III-V semiconductor
alloy GaAs$_{1-x}$N$_x$, where $x$ is the concentration of the nitrogen impurity.
Both GaAs and GaN are technologically important materials in semiconductor industry,
hence there has been considerable interest in alloying GaAs and GaN to obtain optoelectronic properties that bridge nitrides and arsenides.
The simulations are done again with the supercell approach with successively increasing cell size,
containing 16, 32, 64, 128, 256, 512, and finally 1024 atoms.
These correspond to approximate $x$ values of 0.125, 0.063, 0.031, 0.016, 0.008, 0.004, and 0.002 respectively.
Only $\Gamma$-point is used in the simulations of supercells with 512 atoms and more.
For smaller supercells, finite {\bf k}-point meshes are used.
Specifically, they are $1\times 1 \times 2$, $1\times 2 \times 2$, $2\times 2 \times 2$, $2\times 2 \times 4$,
and $2\times 4 \times 4$ {\bf k}-point meshes for system sizes ranging from 256 atoms to 16 atoms, respectively.
The LDA is chosen to be exchange-correlation functional.
The internal geometry of each supercell is fully relaxed.

\subsubsection{Band gap of GaAs$_{1-x}$N$_x$}

\begin{figure}[t]
    \includegraphics[width=0.48\textwidth, clip]{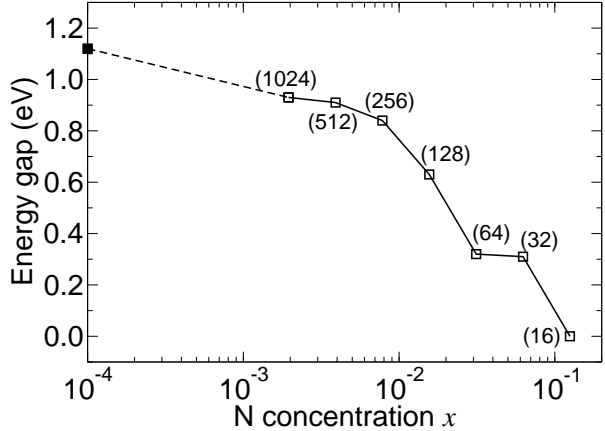}
    \caption{\label{fig:GaAsN_gap}Calculated band gap of GaAs$_{1-x}$N$_x$ as a function of the N concentration $x$.
    The solid square on the left corresponds to the band gap of GaAs bulk.
    The number of atoms used in a supercell is labeled for each blank black point.
    The lines connecting the data points are for guidance.}
\end{figure}

The photoluminescence edge of GaAs$_{1-x}$N$_x$ for small $x$ shows an unexpected redshift,
instead of a blueshift \cite{Weyers/etal:1992} as inferred from the linear interpolation between the
two endpoints (1.4 eV for GaAs and 3.8 eV for GaN).
The narrowing of the alloy band gap $E_\text{g}(x)$ from
the composition-weighted linear average value $\bar{E}_\text{g}(x)=xE_\text{g}(0)+ (1-x)E_\text{g}(1)$ is
called the band-gap bowing, which is a general feature of semiconductor alloys and
can often be described as $\delta E_\text{g}(x) = bx(x-1)$ with $b$ being the optical bowing parameter.
\cite{Bernard/Zunger:1987} In GaAs$_{1-x}$N$_x$, the bowing effect is extremely pronounced with
a large $b$ coefficient of about 16 eV, in stark contrast with most isovalent semiconductor
alloys that have a $b$ value of only a fraction of eV.
This not only leads to a pronounced band-gap narrowing for small $x$ ($x<0.015)$,
but also even a closing of the band gap for large $x$ values.

Such a peculiar behavior has been analyzed in details by several theoretical
studies \cite{Rubio/Cohen:1995,Neugebauer/VandeWalle:1995,Wei/Zunger:1996,Mattila/Wei/Zunger:1999} based on
first-principle calculations, and
the strong bowing effect has been attributed to the substantial lattice mismatch ($>20\%$) between GaAs
and GaN\cite{Rubio/Cohen:1995,Neugebauer/VandeWalle:1995} and the formation of spatially separated and
sharply localized band-edge states. \cite{Wei/Zunger:1996}  However, due to the limitation of the computational
resources, the previous calculations of GaAs$_{1-x}$N$_x$ focused only on the alloy regime
with $x=$0.25, 0.50 and 0.75, \cite{Rubio/Cohen:1995,Neugebauer/VandeWalle:1995,Wei/Zunger:1996} or were
based on empirical pseudopotentials. \cite{Mattila/Wei/Zunger:1999}

The \CodeName~ package with LCAO basis sets allows us to reach large systems with $x$ as low as $0.002$.
The calculated band gaps of GaAs$_{1-x}$N$_x$ systems are shown in Fig.~\ref{fig:GaAsN_gap}.
The blank square points from right to left are the band gaps obtained from
supercell calculations using 16, 32, 64, 128, 256, 512, and 1024 atoms, respectively.
Note that the $x=0$ limit represents the calculated band gap of bulk GaAs, which is 1.13 eV from LDA.
Although LDA based calculation in general underestimates the band gap of materials,
it reproduces very well the experimental finding that
the band gap of GaAs$_{1-x}$N$_x$ gets continuously reduced as $x$ increases.
We note that, the geometries of the supercells in this work are chosen in a special way, {\it i.e.},
by doubling the supercell size successively along the $x$, $y$, and $z$ directions.
Thus the exact behavior of the plot in Fig.~\ref{fig:GaAsN_gap} might slightly changes
if the shape of the supercells is chosen differently.
However, this should not alter the general trend we obtained.
We also find that the closing of the band gap happens when $x$ is larger than 0.125.

\subsubsection{Formation energy of GaAs$_{1-x}$N$_x$}

\begin{figure}[t]
\includegraphics[width=0.48\textwidth,clip]{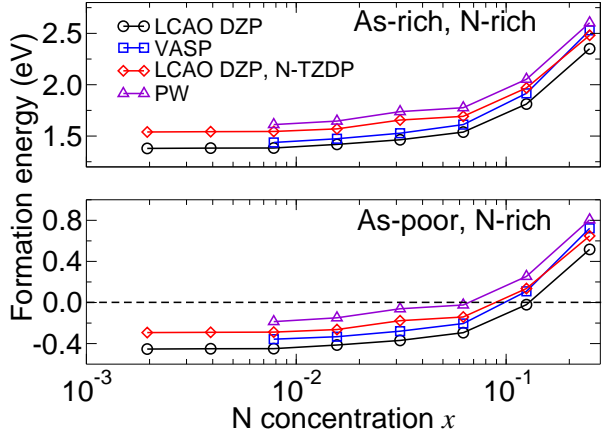}
\caption{\label{fig:GaAsN_FormE}(Color online) The N defect formation energy of GaAs$_{1-x}$N$_x$ as a function of the N
concentration $x$. Two LCAO basis set setups are used in the \CodeName~ calculations: DZP for all elements
(black circles);  DZP for Ga and As, and TZDP for N (red diamonds). Results obtained from VASP (blue squares)
and \CodeName/PW basis (violet triangles) calculations are shown for comparisons.
The energy cutoff is set to 500 eV in VASP calculations.}
\end{figure}

The next relevant issue is the stability of the GaAs$_{1-x}$N$_x$ alloy.
A key quantity here, for small $x$, is the formation energy of an N defect, which is defined as
\begin{equation}
E_f = E(\text{Ga}_n\text{As}_{n-1}\text{N}) - E(\text{Ga}_n\text{As}_n) + \mu(\text{As}) - \mu(\text{N}),
\label{Eq:formE}
\end{equation}
where $E(\text{Ga}_n\text{As}_n)$ is the energy of a Ga$_n$As$_n$ supercell containing $n$ GaAs formula units,
$E(\text{Ga}_n\text{As}_{n-1}\text{N})$ is
the energy of the above supercell but with one As atom replaced by an N impurity atom. $\mu$(As) and $\mu$(N) are the chemical
potentials for the As atom and the N atom, respectively.
The actual value of the atomic chemical potential depends on the chemical conditions in the experiment.

Here we consider two situations: the As-rich $\&$ N-rich condition and the As-poor $\&$ N-rich condition.
In the first one, $\mu(\text{As})=\mu^\text{rich}(\text{As})$ is chosen to be the
energy per atom in the yellow arsenic crystal form.
In the second case, $\mu^\text{poor}(\text{As})$ is given by adding the formation enthalpy of
bulk GaAs to $\mu^\text{rich}(\text{As})$.
The formation enthalpy of bulk GaAs bulk is taken here as the atomization energy of GaAs per atom
at zero temperature (2.00 eV).
Finally, $\mu^\text{rich}(\text{N})$ is chosen to be one half of the total energy of the N$_2$ molecule.
Obviously, the corresponding N concentration in the supercell Ga$_n$As$_{n-1}$N is given by $x=1/n$. We would
like to point out that, by using Eq.~(\ref{Eq:formE}), the N impurities are regarded as point defects, and
the possible interactions between neighboring N defects are neglected.

In Fig.~\ref{fig:GaAsN_FormE}, the formation energy of an N defect is plotted
as a function of the N concentration $x$ for both As-rich (upper panel) and As-poor (lower panel) conditions.
Calculations are done using the \CodeName~ package with two sets of LCAO bases: 1) DZP for
all elements; 2) DZP for Ga and As, and TZDP for N.
The results are compared to those obtained from the \CodeName/PW calculations,
and those obtained by the Vienna Ab-initio Simulation Package (VASP) package
with projected augmented wave method.
\cite{Kresse/Furthmuller:1996a,Kresse/Furthmuller:1996b}
Fig.~\ref{fig:GaAsN_FormE} shows that under the As-rich condition,
the N defect formation energies of GaAs$_{1-x}$N$_x$ are positive for all N compositions,
indicating the very low probability to dope N into GaAs under such condition.
However, under the As-poor condition, the formation energies become
negative for small N concentrations,
which is consistent with the experimental finding that GaAs$_{1-x}$N$_x$
alloy can only be formed for narrow composition range near the
endpoints. \cite{Weyers/etal:1992}

The formation energy curves from \CodeName/DZP calculations follow the same trend as the ones
from \CodeName/PW calculations for both As-rich and As-poor cases in Fig.~\ref{fig:GaAsN_FormE}.
However, there are noticeable differences of calculated formation energies between the two methods.
Specifically, the formation energies associated with the DZP basis set are underestimated by 0.2 to 0.3 eV compared
to the ones from PW calculations.
This is due to the fact that $\mu(N)$ is calculated by half of the N$_2$ energy (cf. Eq.~\ref{Eq:formE}),
and the N$_2$ total energy calculated with the DZP basis still has an appreciable difference from the PW reference result, as shown in Fig~\ref{Fig:N2}.
In fact, by just increasing the N basis set from DZP to TZDP,
the corresponding formation energy difference is largely improved to within 0.1 eV compared to PW results for all cases, as shown in Fig.~\ref{fig:GaAsN_FormE}.
Finally, comparing the \CodeName/PW results and the VASP results reveals
that there exists an appreciable difference between the norm-conserving psedudopotenital treatment and the
projector augmented wave method. Despite these differences,  the general trend of the dependence of the
formation energy on the N concentration is well reproduced within all calculations.

In conclusion, the study of GaAs$_{1-x}$N$_x$ alloy illustrates that the \CodeName~ package with its LCAO basis sets
can be used for reliable defect calculations. We would also like to note that,
for a thorough and faithful treatment of the phase stability problem of the GaAs$_{1-x}$N$_x$ alloy,
one needs to consider the influence of finite temperatures and include the entropy effect,
but this goes beyond the scope of the present work.

%% file: summary.tex
\section{\label{sec:conclusion}Summary}

To summarize, in this paper we introduce a comprehensive first-principles package, named \CodeName~, in
which both plane waves and efficient localized atomic orbitals can be used for electronic-structure calculations.
In particular, we present the mathematical foundation and numerical techniques behind the atomic-orbital-based
implementation within this package.  The performance and reliability of the \CodeName~ package were benchmarked
for a variety of systems containing molecules, solids, surfaces and defects.
Furthermore, we show that the hierarchial atomic basis sets generated with the CGH scheme
allows for a systematic convergence towards the plane-wave accuracy,
and the DZP basis set offers an excellent compromise between accuracy and the computational load,
and can be used in production calculations for most purposes. The package is currently under active development,
with more features and functionalities are being implemented.
With all these efforts, we expect the \CodeName~ package will become a powerful and reliable tool
for simulations of large-scale materials.

%% file: appendix.tex
\section{Appendix}

\subsection{Two-center integrals}
\label{sec:twocenter}
The overlap matrix and the kinetic energy matrix
can be efficiently calculated by two-center integral technique,~\cite{sankey89}
which has been described with full details in Ref. \onlinecite{soler02}.
Here we briefly introduce this algorithm.
The overlap matrix is written as
\begin{equation}
S({\bf R})=\int \phi_{\mu}^*({\bf r})\phi_{\nu}({\bf r}-{\bf R}) d{\bf r}\, ,
\end{equation}
which can be further written as,\cite{soler02}
\begin{equation}
S({\bf R})=\sum_{l=0}^{2l_{max}}\sum_{m=-l}^{l}
S_{l_{\mu}m_{\mu},l_{\nu}m_{\nu},lm}(R)
G_{l_{\mu}m_{\mu},l_{\nu}m_{\nu},lm}
Y_{lm}(\hat{\mathbf{R}})\, .
\label{eq:overlap}
\end{equation}
Here $l_{\mu}$ ($l_{\nu}$) and $m_{\mu}$ ($m_{\nu}$ )are angular momentum and magnetic quantum numbers for orbital $\mu$ ($\nu$).
The radial part is
\begin{equation}
S_{l_{\mu}m_{\mu},l_{\nu}m_{\nu},lm}(R)
=4\pi i^{-l}\int_{0}^{\infty}
j_{l}(kR)f_{\mu}(k)f_{\nu}(k)k^2dk \, .
\end{equation}
Here $f_{\mu}(k)$ and $f_{\nu}(k)$ are one dimensional Fourier transform of
the radial atomic functions introduced in Eq. \ref{eq.radial},
\begin{equation}
f_{\mu}(k)=\sqrt{2\over \pi} (-i)^{l_{\mu}} \int_0^{\infty} r^2 j_{l_{\mu}} (kr)f_{\mu}(r)dr\, .
\end{equation}
$S_{l_{\mu}m_{\mu},l_{\nu}m_{\nu},lm}(R)$ can be tabulated with dense sampling of
distances between $\phi_{\mu}$ and $\phi_{\nu}$.
$G_{l_{\mu}m_{\mu},l_{\nu}m_{\nu},lm}$ in Eq.~\ref{eq:overlap}
is called the Gaunt coefficient,
\begin{equation}
G_{l_{\mu}m_{\mu},l_{\nu}m_{\nu},lm}=\int_{0}^{\pi}
\sin(\theta)d\theta
\int_{0}^{2\pi}
Y_{l_{\mu}m_{\mu}}(\theta,\phi)Y_{l_{\nu}m_{\nu}}(\theta,\phi)
Y_{lm}(\theta,\phi)d\phi,
\end{equation}
which can be calculated and tabulated recursively from the Clebsch-Gordan coefficients. \cite{soler02}
For convenience, the real spherical harmonic functions are actually used in \CodeName.
The overlap matrix formed by an atomic orbital $|\phi_{\mu}\rangle$
and a non-local projector $|\chi_{\alpha lmn}\rangle$ [See Eq. (\ref{eq:nl_pp})]  can be calculated exactly
in the same way.

The kinetic energy operator matrix element,
\begin{equation}
T({\bf R})=\int \phi^*_{\mu}({\bf r})(-\frac{1}{2}\nabla^2)\phi_{\nu}({\bf r}-{\bf R}) d {\bf r}
\end{equation}
is slightly different.
This term can be calculated in a similar way as $S({\bf R})$ by replacing
$S_{l_{\mu}m_{\mu},l_{\nu}m_{\nu},lm}(R)$ with
\begin{equation}
T_{l_{\mu}m_{\mu},l_{\nu}m_{\nu},lm}(R)
=2\pi i^{-l}\int_{0}^{\infty}
j_{l}(kR)f_{\mu}(k)f_{\nu}(k)k^4dk.
\end{equation}
Because $S_{l_{\mu}m_{\mu},l_{\nu}m_{\nu},lm}(R)$, $G_{l_{\mu}m_{\mu},l_{\nu}m_{\nu},lm}$ and
$T_{l_{\mu}m_{\mu},l_{\nu}m_{\nu},lm}(R)$ are all independent of the coordinates of atoms,
they can be tabulated at the beginning of the DFT calculations once for all.
For any given distance between two atoms, the corresponding overlap matrix elements and kinetic energy matrix
elements can be calculated efficiently by interpolation method.

\subsection{Grid-based techniques}
\label{sec:vlocal}
The Hamiltonian matrix elements $V^{loc}_{\mu\nu}$ are evaluated on a uniform real space grid,
with both atomic orbitals and local potentials presented on each grid point.
The local potential is,
\begin{equation}
V^{loc}({\bf r})=V^{L}({\bf r})
+V^{H}({\bf r})+V^{xc}({\bf r}),
\end{equation}
where $V^{L}({\bf r})=\sum_{\bf R}\sum_{\alpha i}v^{L}_{\alpha}({\bf r}-\tau_{\alpha i}-{\bf R})$
is the summation of all the local pseudopotentials for  $i$-th atom of element type $\alpha$.
Plane wave basis and Fast Fourier Transform (FFT) techniques are used here to efficiently evaluate
$V^{L}({\bf r})$ and $V^{H}({\bf r})$ on the grid.
Because the local pseudopotential $v^{L}_{\alpha}({\bf r})$ has a fairly long tail in real space,
it is inefficient to calculate $V^{L}({\bf r})$ directly on a real space grid.
Therefore, $\mathrm{V}^{L}({\bf G})$ is first calculated in reciprocal space as
\begin{equation}
V^{L}({\bf G})=\sum_{\alpha} S_{\alpha}({\bf G}) v^L_{\alpha}({\bf G}) \, ,
\end{equation}
where $S_{\alpha}({\bf G})=\sum_{i}
e^{-i{\bf G}\cdot \tau_{\alpha i}}$
is the structure factor.
An FFT is carried out to bring $V^{L}({\bf G})$ back to real space.
From our tests, this construction processes of $V^{L}(\mathbf{r})$
only take a small portion of total computational time,
even for systems containing thousands of atoms.
This is different from the method used in Ref. \onlinecite{soler02},
where the short-ranged neutral atom potentials are used.
Using the same set of plane wave basis,
the Hartree potential is also first evaluated in reciprocal space
and then be brought back to real space by using an FFT.
The full formula of Hartree potential is
\begin{equation}
V^{H}({\bf r})=4\pi\sum_{{\bf G}\neq 0}
\frac{\rho({\bf G})}{|{\bf G}|^2}
e^{i{\bf G}\cdot{\bf r}}.
\end{equation}

\subsection{Force calculations}
\label{sec:force}
As shown in the main text, the forces evaluated from the basis of atomic orbitals
have four contributions,
\begin{equation}
F=F^{FH} + F^{pulay} + F^{ortho} +F^{Ewald}\, ,
\end{equation}
which are the Feynman-Hellmann force, the Pulay force, the force due to nonorthogonality
of the atomic orbitals, and the Ewald force due to the Coulomb interactions between the ions.
The Ewald force can be calculated analytically using the Ewlad summation techniques, \cite{ewald21}
and therefore not discussed here.
We discuss here the techniques to evaluate the rest three terms as follows.

1.  Feynman-Hellmann force,
\begin{equation}
F^{FH}_{\alpha i}=-\sum_{\bf R} \sum_{\mu\nu} \langle \phi_{\mu{\bf R}} |{\partial H \over \partial
  \tau_{\alpha i}}|\phi_{{\nu}0}\rangle \, .
\end{equation}

In Hamiltonian $H$, only the local and non-local pseudopotentials explicitly depend on the coordinates of ions.
Thus we can further break $F^{FH}_{\alpha i}$ into two terms related to pseudopotentials.
The first one is related to the local pseudopotential, and can be evaluated in reciprocal space,
\begin{equation}
F^{L}_{\alpha i}=
\sum_{{\bf G\neq 0}}i\mathbf{G}e^{-i\mathbf{G}\cdot\tau_{\alpha i}}
V^{L}(\mathbf{G})\rho^{\ast}(\mathbf{G}).
\end{equation}
This method has been shown to be accurate and fast. \cite{nicholas11}
The second term involves the non-local pseudopotential,
\begin{equation}
\begin{split}
F^{NL}_{\alpha i}&=-\sum_{\bf R} \sum_{\mu\nu}\langle \phi_{\mu{\bf R}} |{\partial V^{NL}
\over \partial \tau_{\alpha i}}|\phi_{{\nu}0}\rangle \\
&=-\sum_{\bf R} \sum_{\mu\nu}\sum_{lmn}\Bigl(
\langle\phi_{\mu{\bf R}}|\frac{d\chi_{\alpha ilmn}}{d\tau_{\alpha i}}\rangle
\langle\chi_{\alpha ilmn}|\phi_{\nu 0}\rangle
+\langle\phi_{\mu{\bf R}}|\chi_{\alpha ilmn}\rangle\langle
\frac{d\chi_{\alpha ilmn}}{d\tau_{\alpha i}}|\phi_{\nu 0}\rangle
\Bigr).
\end{split}
\label{eq:FH-nl}
\end{equation}
A non-local pseudopotential projector $\chi_{\alpha ilmn}$ is also a one dimensional numerical orbital like atomic orbital.
Therefore, the above equation can be evaluated efficiently using two-center integral technique introduced before.

2. Pulay force

The existence of Pulay force is due to the fact that the basis set is not complete,
\begin{equation}
F^{pulay}_{\alpha i}=-\sum_{\bf R} \sum_{\mu\nu}\Bigl(
 \langle {\partial \phi_{\mu{\bf R}} \over \partial
  \tau_{\alpha i}} | H|\phi_{\nu0}\rangle
+\langle \phi_{\mu{\bf R}} | H |{\partial \phi_{\nu0} \over \partial
  \tau_{\alpha i}}\rangle \Bigr)\, .
\label{eq:pulay}
\end{equation}
The kinetic energy term and the non-local pseudopotential term in $F^{pulay}_{\alpha i}$ can also be calculated by
the two center integral technique, whereas the local potential term is evaluated by the grid integral technique.
These steps are similar to the calculations of Hamiltonian matrix, except one needs to substitute
$\phi_{\mu}$ and $\phi_{\nu}$ with $\frac{d\phi_{\mu}}{d\tau_{\alpha i}}$ and $\frac{d\phi_{\nu}}{d\tau_{\alpha i}}$.
To evaluate the derivatives of an atomic orbital with respect to
the coordinates of the atoms, one needs to deal with
both radial atomic orbital and spherical harmonic function parts.
The radial part can be calculated numerically on a one dimensional grid
while the second part can be obtained analytically by using the real spherical harmonic functions.

3. The force arises from the fact that atomic orbitals are not orthogonal,
\begin{equation}
 F^{ortho}_{\alpha i} =-\sum_{{\bf R}}\sum_{\mu\nu} E_{\mu\nu}({\bf R}) {\partial S_{\mu\nu}({\bf R}) \over \partial \tau_{\alpha i}}  \,,
\end{equation}
where
\begin{equation}
E_{\mu\nu}({\bf R})
=\frac{1}{N_{k}}\sum_{n\mathbf{k}}f_{n\mathbf{k}}
\epsilon_{n\mathbf{k}}
c_{\mu n,\mathbf{k}}^{\ast}
c_{n\nu,\mathbf{k}}
e^{-i\mathbf{k}\cdot{\bf R}}
\end{equation}
is the element of ``energy density matrix''. \cite{soler02}
The derivative of
overlap matrix with respect to atomic coordinates is,
\begin{equation}
\frac{\partial\mathrm{S}_{\mu\nu}({\bf R})}{\partial\tau_{\nu}}=
-\frac{\partial\mathrm{S}_{\mu\nu}({\bf R})}{\partial\tau_{\mu}}
=\frac{d\mathrm{S}_{\mu\nu}({\bf R})}{d{\bf D}}\,,
\end{equation}
where $\tau_{\mu}$ and $\tau_{\nu}$ are the atomic coordinates
for orbital $\mu$ and $\nu$, and
${\bf D}={\bf R}+\tau_{\nu}- \tau_{\mu}$, is the distance between
two orbitals.
The further expansion of this term is
\begin{equation}
\begin{split}
\frac{d\mathrm{S}_{\mu\nu}({\bf R})}{d\mathbf{D}}
&=\sum_{lm}\frac{d}{dD}
\Bigl[{D^{-l}}{S_{l_{\mu}m_{\mu},l_{\nu}m_{\nu},lm}(D)}
\Bigr]G_{l_{\mu}m_{\mu},l_{\nu}m_{\nu},lm}Y_{lm}(\mathbf{\hat{D}})
D^{l}\mathbf{\hat{D}}\\
&+\sum_{lm}
{D^{-l}}{S_{l_{\mu}m_{\mu},l_{\nu}m_{\nu},lm}(D)}
G_{l_{\mu}m_{\mu},l_{\nu}m_{\nu},lm}\frac{d}{d\mathbf{D}}\Bigl[Y_{lm}(\mathbf{\hat{D}})
D^{l}\Bigr].
\end{split}
\end{equation}
In order to make $Y_{lm}(\mathbf{\hat{D}})$ analytical
at the origin, here it is multiplied by $D^{l}$.
$\frac{d}{dD}
\Bigl({D^{-l}}{S_{l_{\mu}m_{\mu},l_{\nu}m_{\nu},lm}(D)}\Bigr)$
can be calculated numerically using interpolation method
while $\frac{d}{d\mathbf{D}}\Bigl[Y_{lm}(\mathbf{\hat{D}})D^{l}\Bigr]$
can be calculated analytically by using real spherical harmonic functions.%